# Sharp Partial Identification for Survival Model Comparison Without Target Outcomes

Won Gi Choi [a], Sun-Ho Kim [b], Min Soo Kim [c*]

[a] *Department of Mathematics and Statistics, Chonnam National University, Gwangju 61186, Republic of Korea;*

[b] *Graduate School of Data Science, Chonnam National University, Gwangju 61186, Republic of Korea;*

[c] *Department of Statistics, Chonnam National University, Gwangju 61186, Republic of Korea;*

** Corresponding author: Min Soo Kim, Department of Statistics, Chonnam National University, Gwangju 61186, Republic of Korea (kimms@chonnam.ac.kr)*

**Running head:** Partial Identification Without Target Outcomes

## Abstract

Hospitals often need to know whether a new prognostic model outperforms an established one before survival outcomes in their own patients are available. Without these outcomes, the performance difference is only partially identified: it lies in an interval whose width reflects how far the target population may depart from the source. We show that this interval should be built for the difference itself, not by bounding each model's risk separately, because both risks depend on the same unknown outcomes. For the fixed-horizon Brier-risk difference, under bounded shifts in conditional log-odds that may vary in sign and magnitude across patient profiles, we derive the sharp identified set, prove that it is never wider than separate-risk subtraction, characterize exactly when it is narrower, and give a diagnostic, computable without target outcomes, for when a single-shift model narrows the interval only by assuming homogeneity. Censoring-adjusted simultaneous bootstrap envelopes turn these sets into decisions. Across 32 nonzero-sensitivity comparisons in lung cancer, separate-risk bounds were 4.0–11.1 times as wide as direct identified sets, although sampling uncertainty with 35 source deaths prevented directional certification. In a Rotterdam-to-German Breast Cancer Study Group analysis, direct identification supported the candidate model in a comparison that separate subtraction left inconclusive. Hospitals can thus obtain evidence, before local outcomes exist, on whether a candidate model predicts more accurately under explicitly stated population differences.

## 1. Introduction

A hospital considering a positron emission tomography (PET)–radiomics-augmented prognostic model must decide whether it improves on an established clinical model before survival outcomes in its own patients mature. External validation requires attention to differences between development and target populations (Altman and Royston, 2000; Royston and Altman, 2013), yet the target outcomes needed to assess their consequences for predictive performance are unavailable. Our motivating example is two-year survival prediction in non-small-cell lung cancer (NSCLC): models developed at Chonnam National University Hwasun Hospital (CNUHH) are to be compared in Stanford and Veterans Affairs (VA) cohorts from the NSCLC-Radiogenomics collection of The Cancer Imaging Archive, with target outcomes withheld from the comparison.

We compare fixed-horizon Brier risks through the candidate-minus-reference contrast; negative values favor the candidate. Related work addresses comparative predictive accuracy and risk estimation under dataset shift (Diebold and Mariano, 2002; Qiu, Tchetgen and Dobriban, 2024). Source outcomes can inform target performance under conditional transportability, but this assumption cannot be tested without target outcomes (Steingrimsson, 2023; Steingrimsson et al., 2023; Steingrimsson et al., 2024). The most recent of these studies developed a loss-based sensitivity analysis under an exponential tilt, which for a binary outcome shifts the conditional log-odds by a single parameter; applied to a difference of prediction losses, it traces a sensitivity curve for the performance contrast. They identified failure-time outcomes and departures that vary with covariates as open problems. Outcome-free partial identification has also been studied for classification under weak supervision (Maia Polo et al., 2024).

Once departures may differ across patients, many target outcome laws remain plausible, and the performance difference can only be bounded. The natural approach bounds each model's risk and subtracts the intervals. This can be wasteful: both risks are evaluated on the same unknown outcomes, so their extremes need not occur under a single outcome law, and separate bounding can produce intervals wider than the data and assumptions warrant (Manski, 2003). Even when the two predictors are identical, separately bounding their risks can give a contrast interval of positive width, whereas direct identification correctly returns zero. The comparison should instead be bounded directly.

Our contributions are threefold. First, for shifts in the conditional log-odds of survival that are bounded but may vary in sign and magnitude across deployment profiles, we derive the sharp identified set for the Brier-risk contrast, prove that it is never wider than separate-risk subtraction, and characterize exactly when it is strictly narrower; this gain can be computed before any target outcome is observed. Second, we give a first-order diagnostic, also computable without target outcomes, that reveals when a single-shift sensitivity model attains narrower ranges only by assuming homogeneous departures. Third, we develop right-censoring-adjusted estimation and simultaneous confidence envelopes over a prespecified sensitivity grid, which convert identified sets into directional decisions (Figure 1). Across 32 nonzero-sensitivity comparisons in NSCLC, separate-risk bounds were 4.0–11.1 times as wide as the direct identified sets, and in a Rotterdam–GBSG breast-cancer analysis direct identification supported a candidate model where separate subtraction remained inconclusive.

## 2. Methods

### 2.1. Setting and target estimand

Consider a reference survival predictor $q_0(X)$ and an imaging-augmented candidate $q_1(X)$, both developed and locked before evaluation. At a prespecified prediction horizon $\tau$, let $T^*$ be event time and $Y_\tau = \mathbf{1}(T^* > \tau)$ indicate survival beyond that horizon. Here $T$ labels the target population, whereas $T^*$ denotes individual event time. The target population provides patient characteristics and both predictions but not $Y_\tau$. An independent source calibration sample, indexed by $S$, provides right-censored outcomes: $O_i = (X_i, \tilde{T}_i, \Delta_i)$, where $\tilde{T}_i = \min(T_i^*, C_i)$ and $\Delta_i = \mathbf{1}(T_i^* \leq C_i)$. Predictor development is separate from source calibration; the development samples are used only to construct the locked functions. We condition on those functions, and predictor-development uncertainty is not part of the present analysis.

For $j \in \{0,1\}$, we define the target Brier risk using the Brier score, a proper scoring rule for probabilistic prediction (Brier, 1950; Gneiting and Raftery, 2007; Graf et al., 1999):

$$R_T(q_j) = E_T\left[\left(Y_\tau - q_j(X)\right)^2\right].$$

We target the contrast itself,

$$\Delta R_T = R_T(q_1) - R_T(q_0),$$

rather than two marginal risks identified separately. A negative contrast favors the imaging-augmented candidate in expected squared prediction error; a positive value favors the reference. In our NSCLC application, $q_0$ uses clinical variables and $q_1$ adds PET radiomics, with $\tau = 730$ days. Unobserved target outcomes make $\Delta R_T$ generally unidentifiable without restrictions relating source and target outcome laws.

### 2.2. Deployment summary and sensitivity model

We impose restrictions on a prespecified deployment summary $Z = (q_0(X), q_1(X), A)$, where $A$ denotes optional clinical anchors. The principal analyses use $Z = (q_0, q_1)$, without anchors. Write

$$m_S(z) = P_S(Y_\tau = 1 \mid Z = z), \qquad m_T(z) = P_T(Y_\tau = 1 \mid Z = z).$$

Conditioning on $Z$ is sufficient to express the Brier contrast, because both predictions are $Z$-measurable; this does not assert that $Z$ is sufficient to describe survival given the full covariate vector $X$.

Our profile-varying sensitivity model is

$$\text{logit}m_T(z) = \text{logit}m_S(z) + \delta(z), \qquad |\delta(z)| \le \gamma,$$

where $\gamma \ge 0$ is prespecified. Thus $m_T(z)$ belongs to $[m_-(z;\gamma), m_+(z;\gamma)]$, with

$$m_\pm(z;\gamma) = \text{expit}\{\text{logit}m_S(z) \pm \gamma\}.$$

At $\gamma = 0$ we recover conditional transportability at the $Z$ level. This is not implied by transportability given the full covariate vector $X$, because $P(X|Z)$ may differ between populations; $\gamma$ therefore bounds the combined effect of within-profile covariate differences and conditional outcome shift. We work at the Z level because it is low-dimensional, available for every deployed patient, and sufficient for the Brier contrast. At $\gamma = 0.2$ the permissible conditional odds ratio lies between $e^{-0.2} = 0.819$ and $e^{0.2} = 1.221$. The magnitude and direction of the shift can differ across profiles, allowing for heterogeneity potentially arising from treatment practice, outcome recording, or residual case mix. Since $\gamma$ cannot be learned from unlabeled target data, it indexes assumptions rather than an estimated shift. Related sensitivity-bound approaches have been developed for treatment-effect generalization under outcome distribution shift and for causal inference with unmeasured confounding (Asiaee et al., 2026; Dorn, Guo and Kallus, 2025), although their estimands and identifying restrictions differ from those considered here.

**Remark (relation to exponential tilting).** For binary $Y_\tau$, a constant exponential tilt of the conditional law given the same $Z$, with factor $\exp(\eta Y_\tau)$, implies $\delta(z) \equiv \eta$. Hence every such $Z$-conditional tilt with $|\eta| \le \gamma$ belongs to our sensitivity class. This statement does not establish inclusion of the $X$-conditional tilt model studied by Steingrimsson et al. (2024), because coarsening from $X$ to $Z$ changes the conditional laws being compared.

### 2.3. Sharp identification of comparative Brier risk

Define $a(Z) = q_1^2 - q_0^2$ and $b(Z) = q_1 - q_0$, suppressing $X$ in the prediction notation. Since $Y_\tau$ is binary,

$$\Delta R_T = E_T\{a(Z) - 2b(Z)m_T(Z)\}.$$

If the source conditional outcome probability is interior, no finite shift restriction is imposed, and the predictions differ on a target-positive region, both signs of the contrast are compatible with the available data (Proposition 1, Non-identification; proof in Web Appendix A.1). Let $\mathcal{M}_\gamma = \{m: m_-(z;\gamma) \le m(z) \le m_+(z;\gamma)\}$ denote the sensitivity class.

**Theorem 1 (Sharp identified set).** Under the sensitivity model, the identified set of $\Delta R_T$ is

$$\mathcal{I}_\gamma = [L_\gamma, U_\gamma],$$

with

$$L_\gamma = E_T\{a(Z) - 2b(Z)m_L^*(Z)\}, \quad U_\gamma = E_T\{a(Z) - 2b(Z)m_U^*(Z)\}.$$

The endpoint corner assignments are

$$\left(m_L^*(z), m_U^*(z)\right) = \begin{cases} \left(m_+(z;\gamma), m_-(z;\gamma)\right), & b(z) \ge 0, \\ \left(m_-(z;\gamma), m_+(z;\gamma)\right), & b(z) < 0. \end{cases}$$

The extrema are attainable: the endpoint laws obey the sensitivity restriction, and the affine contrast takes every intermediate value along admissible mixtures. Specifically, if $q_1 > q_0$, increasing the conditional survival probability decreases the candidate-minus-reference risk contrast, whereas the opposite corner applies when $q_1 < q_0$. The full proof, including compatibility with a fixed-horizon survival-time law, is given in Web Appendix A.2. As $\gamma$ increases, $L_\gamma$ cannot increase and $U_\gamma$ cannot decrease.

### 2.4. Direct versus separate identification

The conventional alternative first bounds each component risk and subtracts the intervals, giving $\mathcal{I}_\gamma^{\text{sep}} = [L_{1,\gamma} - U_{0,\gamma}, U_{1,\gamma} - L_{0,\gamma}]$.

Each component interval may be sharp, yet its extrema need not be jointly achievable under a single $m_T$.

**Proposition 2 (Direct versus separate identification).** The direct set satisfies

$$\mathcal{I}_\gamma \subseteq \mathcal{I}_\gamma^{\text{sep}}.$$

Writing $d_\gamma(z) = m_+(z;\gamma) - m_-(z;\gamma)$, their widths obey

$$W_{\mathrm{dir}} = 2E_T\big[|q_1 - q_0| d_\gamma(Z)\big],$$

$$W_{\mathrm{sep}} = E_T\big[(|1-2q_1| + |1-2q_0|) d_\gamma(Z)\big], \qquad W_{\mathrm{dir}} \le W_{\mathrm{sep}}.$$

For $\gamma > 0$ the inequality is strict exactly when

$$P_T\Big((q_1 - 1/2)(q_0 - 1/2) > 0,\ d_\gamma(Z) > 0\Big) > 0.$$

Thus the two predictions must fall on the same side of one half on a target-positive region with a nondegenerate sensitivity band. The width difference also has the explicit representation

$$W_{\mathrm{sep}} - W_{\mathrm{dir}} = 2E_T\big[\min(|1-2q_0|, |1-2q_1|)\, \mathbf{1}\{(q_0 - 1/2)(q_1 - 1/2) > 0\}\, d_\gamma(Z)\big].$$

This follows from the elementary identity $|u| + |v| - |u - v| = 2\min(|u|, |v|)$ for $uv > 0$, and zero otherwise, with $u = 1 - 2q_0$ and $v = 1 - 2q_1$. For a profile on which predictions straddle one half, the direct and separate width contributions coincide. By contrast, identical predictions away from one half give $W_{\mathrm{dir}} = 0$ even though $W_{\mathrm{sep}}$ can be positive. Figure 2 illustrates the same-side, straddling, and identical-predictor configurations underlying Proposition 2.

Crucially, $W_{\mathrm{sep}} - W_{\mathrm{dir}}$ depends only on locked target predictions, the target distribution of $Z$, and the source-derived sensitivity limits. Its plug-in value can therefore be assessed without target outcomes, indicating where direct identification is likely to be most informative. When an imaging or biomarker candidate makes predictions close to the clinical reference, the direct width scales with $|q_1 - q_0|$, whereas the separate width need not vanish away from one half. This is an identification-width comparison, not a guarantee that a finite-sample confidence-envelope decision will differ. The proof is in Web Appendix A.3.

**Remark (scalar versus profile-varying shifts)**. Let $w(z) = m_S(z)\{1 - m_S(z)\}$. As $\gamma \to 0$, the width of the comparative-risk range traced by Z-conditional constant tilts with $|\eta| \le \gamma$, relative to $W_{dir}$, converges to $\kappa = |E_T\{b(Z)w(Z)\}|/E_T\{|b(Z)|w(Z)\} \in [0,1]$ (Web Appendix A.6). Thus $\kappa = 1$ when $q_1 - q_0$ has a constant sign, whereas $\kappa < 1$ when the sign is mixed and approaches zero when the w-weighted contributions of the two sign groups nearly cancel, because a common shift moves them in opposite directions. Like $W_{sep} - W_{dir}$, $\kappa$ is computable from source calibration and unlabeled target predictions; it indicates how strongly conclusions from a scalar analysis depend on homogeneity of the departure.

**2.5. Estimation with right-censored source outcomes**

We estimate $m_S(z)$ from the source calibration sample using inverse-probability-of-censoring weighting (IPCW) (Bang and Tsiatis, 2000; Gerds and Schumacher, 2006). Let $H = (Z, V_C)$ contain prespecified censoring covariates and $G(t \mid H) = P_S(C > t \mid H)$. We assume $T^* \perp C \mid H$ and positive censoring survival through $\tau$. The primary analyses set $H = Z = (q_0, q_1)$ and fit an unpenalized conditional Cox censoring model with a Breslow baseline (Breslow, 1972; Cox, 1972). For source individual $i$, the fixed-horizon weight is

$$\widehat{\omega}_i = \frac{\mathbf{1}(\tilde{T}_i > \tau)}{\hat{G}(\tau \mid H_i)} + \frac{\mathbf{1}(\tilde{T}_i \leq \tau, \Delta_i = 1)}{\hat{G}(\tilde{T}_i \mid H_i)}.$$

Patients censored before $\tau$ receive zero weight. An unpenalized IPCW-weighted logistic regression estimates

$$m_S(z; \theta, \xi) = \operatorname{expit}\{\theta^\top \phi(z; \xi)\},$$

where $\phi(z; \xi)$ contains an intercept and the locked prediction scores standardized by source-estimated parameters $\xi$. For each $\gamma$, estimated $m_\pm$ yield plug-in $\hat{L}_\gamma$ and $\hat{U}_\gamma$ by the corner assignments in Theorem 1 and averaging over target observations. Because $q_0$ and $q_1$ are locked, the sign of $b(z)$ that selects each corner is fixed, not estimated.

The implementation requires (i) censoring positivity, $\inf_{t \leq \tau} G(t \mid H) \geq \epsilon_G > 0$ on the relevant support; (ii) bounded-density-ratio transport positivity, $P_T^Z \ll P_S^Z$ with $dP_T^Z/dP_S^Z \leq B_0 < \infty$; and (iii) source outcome interiority, $\epsilon_m \leq m_S(z) \leq 1 - \epsilon_m$. These are population assumptions, not properties certified by empirical diagnostics. Estimated censoring probabilities are floored at 0.02 as a numerical safeguard, assumed asymptotically inactive. In the reported NSCLC and Rotterdam–GBSG point estimates and bootstrap refits, this safeguard was not activated, and nuisance-model fitting failures were not observed. Neither numerical stabilization nor agreement between different censoring estimators (Web Table 2) establishes censoring-model robustness. Full fitting specifications and diagnostics belong to Web Appendix B.

**2.6. Simultaneous inference across the sensitivity grid**

Inference for partially identified parameters and identified sets has been studied extensively (Chernozhukov, Hong and Tamer, 2007; Imbens and Manski, 2004; Stoye, 2009). Let $\mathcal{G} = \{\gamma_1, \dots, \gamma_K\}$ be a finite, prespecified sensitivity grid. We use a joint nonparametric pairs bootstrap: source calibration and target observations are resampled independently; within each source resample, censoring, standardization, and outcome nuisance models are refitted; the complete collection $(\hat{L}_\gamma, \hat{U}_\gamma)_{\gamma \in \mathcal{G}}$ is recalculated. Locked predictors remain fixed. The dependence across sensitivity values is retained within each bootstrap replicate, which is essential because all grid-indexed bounds reuse the same source and target observations. Let $\hat{s}_{L,\gamma}, \hat{s}_{U,\gamma}$ denote bootstrap standard deviations. In bootstrap replicate $r$, form

$$\widehat{M}^{*r} = \max_{\gamma \in \mathcal{G}} \max \left\{ \frac{\left|\hat{L}_\gamma^{*r} - \hat{L}_\gamma\right|}{\hat{s}_{L,\gamma}}, \frac{\left|\widehat{U}_\gamma^{*r} - \widehat{U}_\gamma\right|}{\hat{s}_{U,\gamma}} \right\}.$$

Let $\hat{c}_{1-\alpha}$ be its empirical $(1-\alpha)$ quantile. The outer simultaneous confidence envelope is

$$\underline{L}_\gamma = \hat{L}_\gamma - \hat{c}_{1-\alpha}\hat{s}_{L,\gamma}, \qquad \overline{U}_\gamma = \widehat{U}_\gamma + \hat{c}_{1-\alpha}\hat{s}_{U,\gamma}.$$

For the separate-risk comparator, the component bounds are recomputed in identical resamples, with their own maximum-statistic critical value.

**Theorem 2 (Simultaneous inference for the identified-set endpoints).** Suppose **(A1)** both predictors are locked; **(A2)** the source outcome and conditional censoring models are correctly specified, and their estimators satisfy the required asymptotic linearity and regularity conditions; **(A3)** censoring positivity, interiority, and bounded-density-ratio transport positivity hold, with numerical truncation asymptotically inactive; and **(A4)** endpoint differentiability and positive marginal-variance conditions hold. Conditions A1–A4 summarize conditions R1–R11 in Web Appendix A.5. Then, as $n_S, n_T \to \infty$ with $n_T/(n_S + n_T) \to \rho \in (0,1)$, the endpoint estimators are jointly asymptotically linear, the joint pairs bootstrap consistently estimates the studentized maximum distribution, and

$$P\{\underline{L}_\gamma \leq L_\gamma,\ U_\gamma \leq \overline{U}_\gamma \text{ for every } \gamma \in \mathcal{G}\} \geq 1 - \alpha + o(1).$$

This is simultaneous outer-envelope coverage of the identified sets, not merely containment of one true contrast at a selected $\gamma$. The statement concerns an ideal bootstrap law, with a finite-replicate critical value approximating its quantile, and it does not ensure nominal coverage at small sample sizes.

### 2.7. Sensitivity interpretation and decision summaries

For candidate superiority, define the population robustness value

$$\gamma^*_{\text{cand}} = \sup\{\gamma \geq 0 : U_\gamma < 0\},$$

and analogously

$$\gamma^*_{\text{ref}} = \sup\{\gamma \geq 0 : L_\gamma > 0\}$$

(Proposition 3; proof in Web Appendix A.4). Empty sets receive value zero; where the sign persists to the examined upper bound, we report that the value exceeds the range. Because $U_\gamma$ is nondecreasing and $L_\gamma$ is nonincreasing, they describe the furthest sensitivity allowance within which the identified contrast retains a given sign. Robustness values summarize population identification, whereas sampling-aware decisions use the confidence envelopes. At any evaluated $\gamma$, we label the comparison CANDIDATE SUPPORTED if $\overline{U}_\gamma < 0$, REFERENCE SUPPORTED if $\underline{L}_\gamma > 0$, and DEFER otherwise. DEFER denotes insufficient directional evidence, not equivalence; none of the labels alone recommends clinical deployment. Decision-making under partial identification has also been studied in a broader decision-theoretic framework (Christensen, Moon and Schorfheide, 2026); the present labels instead summarize confidence-envelope evidence without optimizing a clinical utility function.

The prespecified grid is $\mathcal{G} = \{0, 0.1, 0.2, 0.3, 0.5\}$, corresponding to maximum odds multipliers of 1.00, 1.11, 1.22, 1.35, and 1.65. For example, $\gamma = 0.2$ allows the conditional odds of two-year survival among target patients with identical predictions to differ from those in the source by a factor between 0.82 and 1.22. The grid is a prespecified illustrative range rather than a clinically validated one; we therefore treat the full sensitivity curve and the robustness values, not any single $\gamma$, as the primary summary. When an external estimate $\mu$ of the target marginal survival probability is available, for example from a registry, any $\gamma$ for which $\mu$ lies outside $[E_T m_-(Z;\gamma), E_T m_+(Z;\gamma)]$ is incompatible with that estimate, which gives a lower bound on plausible $\gamma$ analogous to the marginal anchoring of Steingrimsson et al. (2024); if $\mu$ is itself estimated, $\gamma$ should be ruled out only when this interval is disjoint from a confidence interval for $\mu$. Target outcomes cannot be used to select $\gamma$ retrospectively. An identified set that excludes zero does not imply the simultaneous confidence envelope must also exclude zero.

## 3. Simulation Study

### 3.1. Objectives and designs

We evaluated (i) the geometry of direct versus separate identification, (ii) finite-sample calibration of the simultaneous confidence envelopes, (iii) directional decisions under distinct signal regimes, and (iv) sensitivity to censoring, source information, overlap, and outcome-model misspecification. All studies used locked predictors; simulated target outcomes determined the known population truth but were not supplied to the evaluation procedure. The simulation objectives, designs, and performance measures were organized following general guidance for simulation studies of statistical methods (Morris, White and Crowther, 2019). Full data-generating mechanisms, prespecified grids, and numerical settings appear in Web Appendix C and Web Tables 3–8.

Four designs were examined: (i) a geometry illustration varying same-side prediction profiles; (ii) an application-scale comparator with 114 source and 60 or 80 target observations and marginal Kaplan–Meier IPCW (Kaplan and Meier, 1958); (iii) primary-procedure simulations matching NSCLC and Rotterdam–GBSG information scales using conditional Cox–Breslow censoring and IPCW-weighted logistic estimation with joint source–target bootstrap inference; and (iv) a five-regime decision study with 3,000 source and 1,500 target observations, marginal or stratified Kaplan–Meier censoring estimation, and 999 bootstrap replicates. Each Monte Carlo condition used 1,000 replications; the geometry illustration used a separate large numerical sample.

### 3.2. Performance measures

Identification performance was measured by direct and separate set widths. The primary inferential measure was simultaneous outer-envelope coverage: at every sensitivity-grid value, the lower confidence limit must not exceed the true lower endpoint and the upper confidence limit must not fall below the true upper endpoint (Theorem 2). We also evaluated two-sided simultaneous endpoint coverage, distinct from single-contrast containment in the marginal Kaplan–Meier comparator. Monte Carlo coverage intervals used Wilson's binomial method (Wilson, 1927); zero wrong-direction decisions were summarized by exact one-sided Clopper–Pearson upper limits (Clopper and Pearson, 1934).

### 3.3. Results

Table 1 reports finite-sample coverage for the primary conditional Cox–Breslow procedure. With 114 source calibration patients and approximately 65% censoring, outer-envelope coverage was 0.861 and 0.847 for target sizes 60 and 80, respectively, materially below the nominal 0.95. At the Rotterdam–GBSG information scale (516 source calibration patients, 686 target patients, approximately 38% censoring), coverage was 0.946, with a Monte Carlo confidence interval including 0.95. Two-sided endpoint coverage was similar. About 59% of these source patients were censored before $\tau$, versus 38–43% in the NSCLC calibration sample (Web Table 8).

In the application-scale marginal Kaplan–Meier comparator at $\gamma = 0.2$, separate-to-direct width ratios were 3.43–3.63 for the identified sets and 1.47–1.59 for the simultaneous confidence envelopes. The controlled-geometry exercise additionally confirmed that the reduction depends on the arrangement of locked predictions, rather than on a universal multiplicative improvement (Web Appendix C; Web Figure 1).

Two-sided grid-wide endpoint coverage in the decision-regime study was 0.939–0.950, and the ambiguous regime always yielded DEFER. In candidate-superior mirror scenarios of the application-scale comparator, no wrong-direction REFERENCE SUPPORTED decision occurred in 1,000 replications per condition (exact one-sided 95% upper bound, approximately 0.003). In a 15-cell factorial (Web Table 8; Web Figure 3), coverage generally rose with source size and fell with censoring; among the cells examined it was closest to nominal with at least 300 source patients and at most 50% censoring, but stayed at most 0.92 at 65% censoring even with 516 patients. Cells with similar numbers of source events (about 180) differed in coverage (0.920 versus 0.947), so event count alone did not determine coverage. Additional diagnostics of the 114-patient, 65%-censoring factorial cell indicated upward endpoint bias and underestimated bootstrap standard errors (Web Appendix C.4).

Overlap and outcome-model misspecification effects appear in Web Appendix C (Web Tables 5 and 6). No calibrated finite-sample correction was evaluated.

# 4. NSCLC Application

## 4.1. Clinical question and study populations

We compared a PET-radiomics-augmented prognostic model with a clinical model for two-year overall survival before external target outcomes were available. Two predictors developed at CNUHH were evaluated in Stanford and Veterans Affairs (VA) cohorts from the public NSCLC-Radiogenomics collection of The Cancer Imaging Archive (Bakr et al., 2018). These target data are publicly available in de-identified form, and no additional participants were enrolled. Target outcomes were withheld from the finalized outcome-free procedure. Because these cohorts were used in our antecedent model-development work (Choi, Kim and Kim, 2026a), the application is a retrospective emulation rather than a prospectively blinded validation.

The CNUHH source data comprised 400 patients with lung cancer whose tumor regions of interest were manually delineated by a nuclear medicine physician; 179 of these patients were also included in previously reported cohorts (Oh et al., 2021; Oh et al., 2023). We excluded 57 patients with small-cell lung cancer and one with a non-positive recorded survival time, leaving 342 patients with NSCLC and 105 observed deaths. The CNUHH study was approved by the Institutional Review Board of Chonnam National University Hwasun Hospital (CNUHH-2019-194), which waived informed consent because of the retrospective design. For each of four prespecified random seeds, an event-stratified two-to-one split produced a predictor-development sample of 228 patients (70 deaths) and a nonoverlapping source-calibration sample of 114 (35 deaths). The seeds defined distinct locked configurations, not independent replications.

Target samples comprised 60 Stanford patients (22 observed deaths) and 80 VA patients (32 observed deaths; Web Table 9). In VA, seven of 88 initially available cases lacked sufficient harmonized metabolic tumor volume for feature extraction and one had failed PET-mask registration. CNUHH tumors were manually delineated, whereas target tumors were segmented using nnU-Net (Isensee et al., 2019); a common 40%-of-maximum-standardized-uptake-value threshold defined metabolic tumor volume for radiomic extraction. Imaging and segmentation differences could nevertheless remain.

### 4.2. Locked predictors and outcome-free evaluation

The reference, $q_0$, was a clinical-only Cox model using age, sex, ever-smoking status, log-transformed pack-years, and clinical T/N/M stage. The candidate, $q_1$, added PET radiomics: 1,112 initial features underwent source-only screening and sparse principal component analysis (Zou, Hastie and Tibshirani, 2006), followed by ridge-penalized Cox fitting with the clinical predictor as an offset. Screening, tuning, and fitting used only CNUHH development data.

The 730-day horizon was prespecified based on clinical relevance and source follow-up. The deployment summary was $Z = (q_0, q_1)$. Nuisance models and inference were as in Sections 2.5–2.7, using the 114-patient calibration subset and 999 bootstrap replicates. All settings were fixed before evaluation.

### 4.3. Identification, uncertainty, and decisions

Across the two target cohorts, four locked configurations, and five sensitivity values, all 40 prespecified evaluations yielded DEFER: none of the simultaneous confidence envelopes excluded zero. In the illustrative seed-42 configuration at $\gamma = 0.2$ (Figure 3), the direct identified sets were [+0.0021, +0.0147] in Stanford and [+0.0248, +0.0410] in VA, both excluding zero in favor of the reference at the identification level. Nevertheless, the corresponding direct simultaneous envelopes, [−0.0237, +0.0455] and [−0.0267, +0.0958], crossed zero. Thus, identification resolved the sign in these two cases, but sampling uncertainty prevented certification. Because every envelope contained zero, any outward enlargement would retain these DEFER decisions; this does not, however, remedy the undercoverage documented in Section 3.

Direct identification reduced identification width substantially relative to separate-risk subtraction: the separate-to-direct width ratio was 4.53 in Stanford and 4.02 in VA. Across all 32 configuration-by-sensitivity cells with $\gamma > 0$, this ratio ranged from 4.0 to 11.1. Corresponding confidence-envelope width ratios were smaller, 1.84 and 1.48, reflecting sampling uncertainty. Web Tables 10–12 report seed-specific specifications, full results, and prespecified one- and three-year analyses (all DEFER).

In the prediction-profile space, source–target classification yielded AUCs of 0.728–0.783; at most 1.67% of target profiles lay outside the source rectangular range, and importance-weight effective sample size was 87.9%–92.6% of target sample size. These diagnostics (Web Table 13) indicate distributional differences but cannot verify positivity.

### 4.4. Retrospective context and clinical interpretation

After the outcome-free decisions had been fixed, target outcomes were examined only in a separate retrospective descriptive analysis. Averaged over four configurations, Stanford had a candidate-minus-reference Brier-risk contrast of −0.0076 and C-index (Harrell Jr, Lee and Mark, 1996) contrast of +0.027; in VA, the corresponding contrasts were +0.0237 and −0.012. Point estimates differed across cohorts; seed-specific intervals generally included zero (Web Table 14).

An outcome-definition limitation remains. Overall survival in CNUHH was indexed from clinical diagnosis (Oh et al., 2021; Oh et al., 2023), whereas the Stanford and VA outcomes were indexed from the CT scan date, closer to the intended imaging-based prediction occasion. Although antecedent CNUHH work placed PET imaging near diagnosis, the diagnosis-to-imaging interval could not be quantified for these 342 patients. Different time origins may alter conditional survival probabilities; the analysis accommodates such differences only to the extent that they fall within the assumed $\gamma$ bound, and does not reconcile the two outcome definitions.

## 5. Rotterdam–GBSG Application

### 5.1. Clinical question and analysis design

We used node-positive patients from the publicly available, de-identified Rotterdam and GBSG breast-cancer cohorts distributed with the survival package in R (Royston and Altman, 2013; Sauerbrei et al., 1999); no additional participants were enrolled. The practical question is whether a prognostic model incorporating additional indicators of nodal disease burden and hormone-receptor status has lower prediction error in the GBSG population than a clinical reference model, before GBSG recurrence outcomes are supplied to the comparison procedure. This is a retrospective demonstration using a public benchmark, not prospective clinical validation.

The Rotterdam node-positive subset (as in Royston and Altman, 2013) was the source ($n = 1{,}546$), with nonoverlapping predictor-development ($n = 1{,}030$) and calibration ($n = 516$) subsets; GBSG was the target ($n = 686$). The prespecified outcome was five-year recurrence-free survival ($\tau = 1{,}826$ days), with follow-up administratively truncated at seven years. The reference Cox predictor included age, menopausal status, tumor size, and hormonal treatment. The candidate added estrogen-receptor (ER) status and transformed positive lymph-node count ($nodes^{-1/2}$). These covariates were selected on established prognostic grounds, not after examining GBSG outcomes. Four prespecified seeds gave four distinct locked predictor configurations, rather than independent replication trials or components of a voting rule.

For both predictors, the deployment summary was $Z = (q_0, q_1)$. Estimation and inference were as for NSCLC, with $B = 999$ bootstrap replicates. Only source outcomes and unlabeled target profiles entered the outcome-free comparisons; GBSG outcomes were reserved for a separate descriptive assessment.

### 5.2. Direct and separate decisions

Table 2 reports both decisions for every one of the 20 prespecified seed-by-sensitivity comparisons. The procedures agreed in 19 cases: at $\gamma = 0$, both methods returned CANDIDATE SUPPORTED in all four configurations; at $\gamma = 0.1$both methods did so for seeds 42 and 62; and at $\gamma \geq 0.2$, except for the discordant seed-62 configuration at $\gamma = 0.2$ the procedures returned DEFER. At the identification level, the estimated robustness value for candidate superiority was 0.29–0.47 under direct identification but 0.15–0.27 under separate subtraction in all four configurations; after sampling uncertainty was incorporated, the decision differed in one cell.

In that differing configuration, the direct simultaneous envelope was $[-0.0486, -0.0017]$, excluding zero and yielding CANDIDATE SUPPORTED. The separate-risk envelope was $[-0.0586, +0.0096]$, which included zero and yielded DEFER. The corresponding plug-in identified sets were $[-0.0365, -0.0156]$ and $[-0.0463, -0.0058]$. Because the procedures used identical bootstrap resamples, nuisance fits, and sensitivity values, differing only in endpoint construction and the resulting method-specific critical values (Section 2.6), the discordance reflects whether the shared target outcome law was preserved. Figure 4 displays the identified sets and simultaneous confidence envelopes for this discordant configuration; numerical endpoints for all 20 comparisons are reported in Web Table 15. This cell shows that direct identification can change a statistical decision; it is not a confirmatory test of candidate superiority across the four configurations.

At this information scale, simulated outer coverage was 0.946 (Table 1), which does not validate the GBSG outcome-shift assumption or the seed-62 certification.

For comparison, we traced the contrast under Z-conditional constant tilts with $|\eta| \leq \gamma$, the binary-outcome analogue of the model of Steingrimsson et al. (2024) (Web Table 1). The candidate predicted higher survival than the reference for 46%–47% of GBSG profiles, so a common shift moved the two groups' contributions in opposite directions; κ ranged from 0.05 to 0.09, and at $\gamma = 0.2$ the scalar range was 5%–9% as wide as the direct identified set. Correspondingly, the population robustness value for candidate superiority was 1.39–1.92 under constant tilts but 0.29–0.47 under the profile-varying class. In NSCLC, $P_T(q_1 > q_0)$ ranged from 0.60 to 0.95 across the four locked configurations and two cohorts, and the width ratio ranged from 0.08 to 0.95; $\kappa$ differed from the width ratio by at most 0.002 in every setting (Web Table 1). Narrower scalar ranges therefore reflect an additional homogeneity assumption rather than additional information, and the certified robustness of a scalar analysis can depend heavily on that assumption.

A separate retrospective GBSG outcome analysis gave an average candidate-minus-reference Brier-risk contrast of $-0.0097$ (bootstrap interval containing zero) and C-index contrast of $+0.081$ (interval excluding zero). Those outcomes neither selected the candidate variables nor determined the outcome-free decisions.

## 6. Discussion

When target outcomes are unavailable, comparing prediction models requires sharp bounds on their performance difference rather than the unnecessarily wide intervals obtained by bounding their risks separately. Because both risks share the same unidentified outcome law, the direct identified set is sharp and never wider than separate-risk subtraction; across 32 nonzero-sensitivity comparisons in lung cancer, separate-risk bounds were 4.0–11.1 times as wide. The procedure suggests a simple workflow: compute the width gain and κ (Section 2.4) from source calibration data and unlabeled target predictions, report the identified set across the sensitivity range together with robustness values, and certify a direction only when the simultaneous envelope excludes zero. In NSCLC, with only 35 observed source deaths, sampling uncertainty prevented directional certification even in comparisons whose identified sets excluded zero; DEFER therefore signaled that more outcome maturation or a larger calibration cohort was needed, not that the models performed equally. In Rotterdam–GBSG, preserving the shared outcome law changed one decision from inconclusive to candidate supported.

Relative to the exponential-tilt analysis of Steingrimsson et al. (2024), which fixes one outcome law per sensitivity parameter for fully observed outcomes and is defined conditional on the full covariate vector, our class is defined on the deployment summary, admits profile-varying departures, and accommodates right-censored source outcomes. Constant tilts given $Z$ are a restriction of this class, and κ, a first-order $(\gamma \to 0)$ width ratio, indicates before target outcomes are observed whether the two formulations will differ materially; in Rotterdam–GBSG the constant-shift range at $\gamma = 0.2$ was only 5%–9% as wide, reflecting its homogeneity assumption rather than additional information. Neither the labels nor the Brier contrast addresses clinical utility, patient harm, cost-effectiveness, or implementation feasibility, and the opposite retrospective point estimates in Stanford and VA caution against unconditional model transfer.

The sensitivity bound cannot be learned from unlabeled target data, and overlap diagnostics do not verify it. The identification results do not depend on working models, but the reported inference does: the primary plug-in estimator requires correctly specified source outcome and censoring models. Under the rate and density-ratio conditions of Proposition 4, a supplementary one-step estimator removes the leading plug-in bias due to outcome-regression error; it is not doubly robust in the model-specification sense, did not improve NSCLC-scale coverage, and left all application decisions unchanged (Web Appendix A.7; Web Tables 16 and 17), so the prespecified plug-in estimator was retained. Unlike the identification results, directional certification relies on the asymptotic envelopes of Theorem 2, whose simulated coverage approached nominal only with moderate-to-large source samples and at most 50% censoring; envelopes from small source samples, such as our NSCLC application, should therefore be read as approximate. Because source and target survival were indexed from diagnosis and from CT scan, respectively (Section 4.4), the NSCLC analysis is not a definitive external validation from a common time origin. Future work should address nuisance-robust inference, finite-sample calibration of the envelope critical value, and alignment of outcome definitions.

**Acknowledgements**

The authors acknowledge The Cancer Imaging Archive (TCIA) and the contributors to the NSCLC-Radiogenomics collection for making the data available. ChatGPT (OpenAI) was used to assist with English-language editing and stylistic refinement. The authors retain full responsibility for the content, interpretations, and conclusions of this manuscript.

**Funding**

This work was supported by the Global-Learning & Academic research institution for Master's·PhD students, and Postdocs (LAMP) Program of the National Research Foundation of Korea (NRF), funded by the Ministry of Education (No. RS-2024-00442775).

**Supplementary Material**

Web Appendices A–C, Web Tables 1–17, and Web Figures 1–3 referenced in Sections 2–5 follow the main text in this preprint. Analysis code is available at https://github.com/wongi-hash/brier-contrast-partial-identification.

**Conflict of interest**

None declared

**Data Availability Statement**

Code, public-data analysis pipelines, synthetic verification fixtures, and aggregate results are available at https://github.com/wongi-hash/brier-contrast-partial-identification and on Zenodo as version 1.2.3 (Choi, Kim and Kim, 2026b). The code is released under the MIT License. Restricted institutional NSCLC data and derived prediction or calibration inputs cannot be publicly redistributed; consequently, the public archive supports auditing through aggregate outputs but not complete rerunning of the institutional analyses.

**Table 1.** Targeted finite-sample performance of the primary conditional Cox–Breslow procedure

| Information scale | Source $n_S$ | Target $n_T$ | Mean censoring | Outer coverage (95% MC CI) | Two-sided endpoint coverage (95% MC CI) |
|---|---|---|---|---|---|
| NSCLC | 114 | 60 | 0.651 | 0.861 [0.838, 0.881] | 0.859 [0.836, 0.879] |
| NSCLC | 114 | 80 | 0.649 | 0.847 [0.823, 0.868] | 0.847 [0.823, 0.868] |
| Rotterdam–GBSG | 516 | 686 | 0.377 | 0.946 [0.930, 0.958] | 0.943 [0.927, 0.956] |

Notes: Coverage is simultaneous over the sensitivity grid, with nominal level 0.95; MC CI denotes the 95% Monte Carlo confidence interval for the estimated coverage probability. The table uses the primary conditional Cox–Breslow implementation, not the marginal Kaplan–Meier comparator. Each setting used 1,000 Monte Carlo replications. Full simulation details are provided in Web Appendix C.

**Table 2.** Rotterdam-to-GBSG direct/separate decisions across all prespecified locked configurations and sensitivity values

| Seed | $\gamma = 0$ | 0.1 | 0.2 | 0.3 | 0.5 |
|---|---|---|---|---|---|
| 42 | CAND/CAND | CAND/CAND | DEFER/DEFER | DEFER/DEFER | DEFER/DEFER |
| 52 | CAND/CAND | DEFER/DEFER | DEFER/DEFER | DEFER/DEFER | DEFER/DEFER |
| 62 | CAND/CAND | CAND/CAND | CAND/DEFER* | DEFER/DEFER | DEFER/DEFER |
| 72 | CAND/CAND | DEFER/DEFER | DEFER/DEFER | DEFER/DEFER | DEFER/DEFER |

Notes: Entries show direct/separate decisions. CAND denotes CANDIDATE SUPPORTED. The asterisk marks the sole discordance (seed 62, $\gamma = 0.2$). Decisions rely on method-specific simultaneous confidence envelopes, not the plug-in identified sets. The four seeds represent distinct locked predictor configurations, not independent replications or voting components. Complete numerical intervals appear in Web Table 15. Simultaneous coverage holds over the sensitivity grid within each locked configuration; no multiplicity adjustment is made across the four configurations.

**Figure 1.** Conceptual overview of outcome-free comparison of two locked survival predictors by direct partial identification and simultaneous inference.

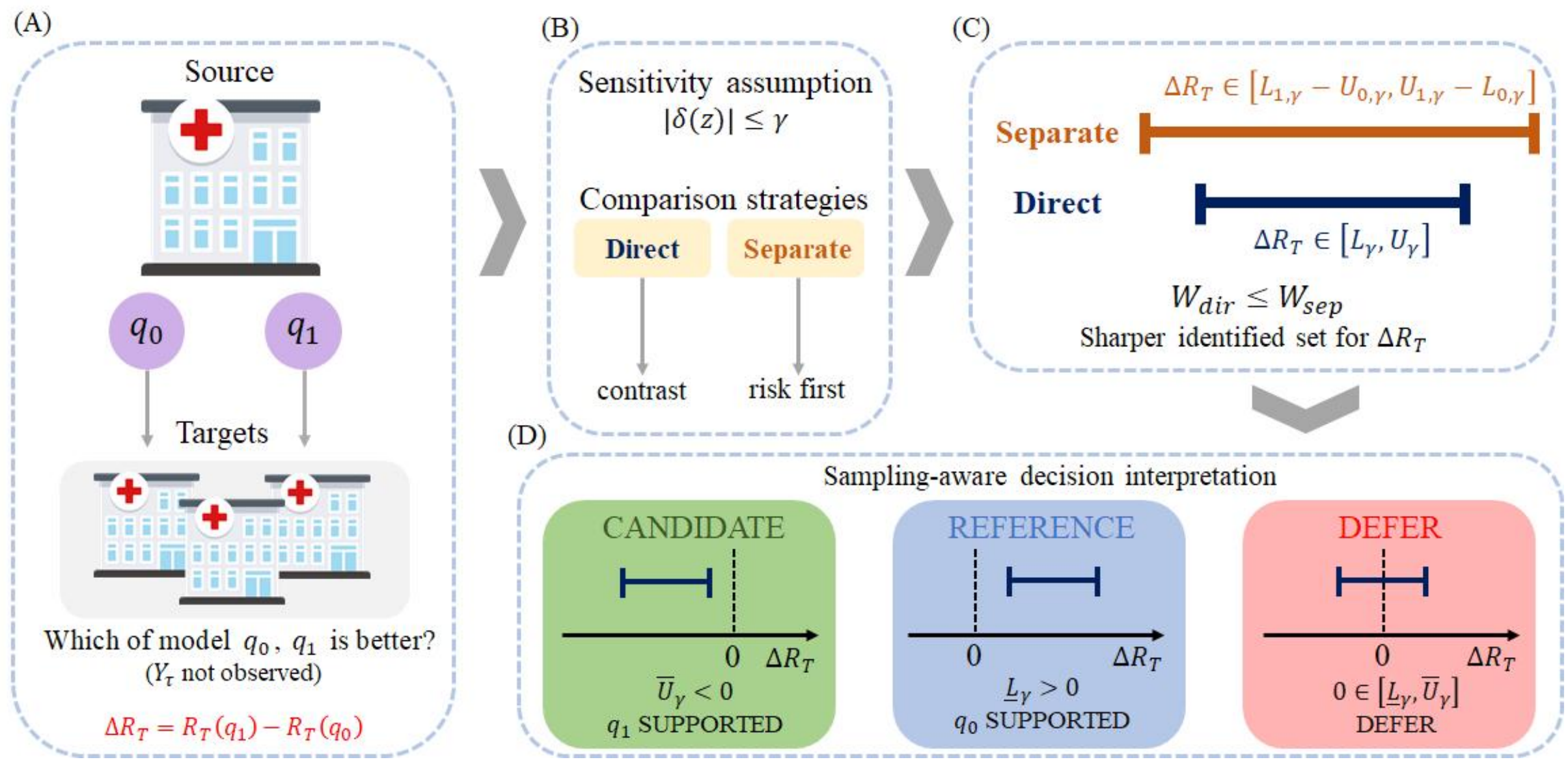

(A) A locked reference predictor $q_0$ and candidate predictor $q_1$ are compared in target populations for which the fixed-horizon outcome $Y_\tau$ is unavailable. (B) A bounded conditional log-odds shift, $|\delta(z)| \leq \gamma$, defines the sensitivity class; the Brier-risk contrast may be identified directly or by separately bounding the component risks. (C) Direct identification preserves the shared unidentified target outcome law and yields a sharp identified set $[L_\gamma, U_\gamma]$ that is no wider than the interval obtained by separate-risk subtraction. (D) Sampling uncertainty is incorporated through simultaneous confidence envelopes for the identified-set endpoints. CANDIDATE SUPPORTED is assigned when $\overline{U}_\gamma < 0$, REFERENCE SUPPORTED when $\underline{L}_\gamma > 0$, and DEFER otherwise.

*Alt text:* Four-panel workflow diagram. Panel A shows two locked survival predictors evaluated in target hospitals without observed survival outcomes. Panel B introduces the sensitivity model and direct versus separate identification. Panel C illustrates that the direct identified set is contained within the separate interval. Panel D shows how simultaneous confidence envelopes relative to zero determine CANDIDATE SUPPORTED, REFERENCE SUPPORTED, or DEFER.

**Figure 2.** Direct versus separate identified sets for the target Brier-risk contrast $\Delta R_T$ *at* $\gamma =$ 0.2.

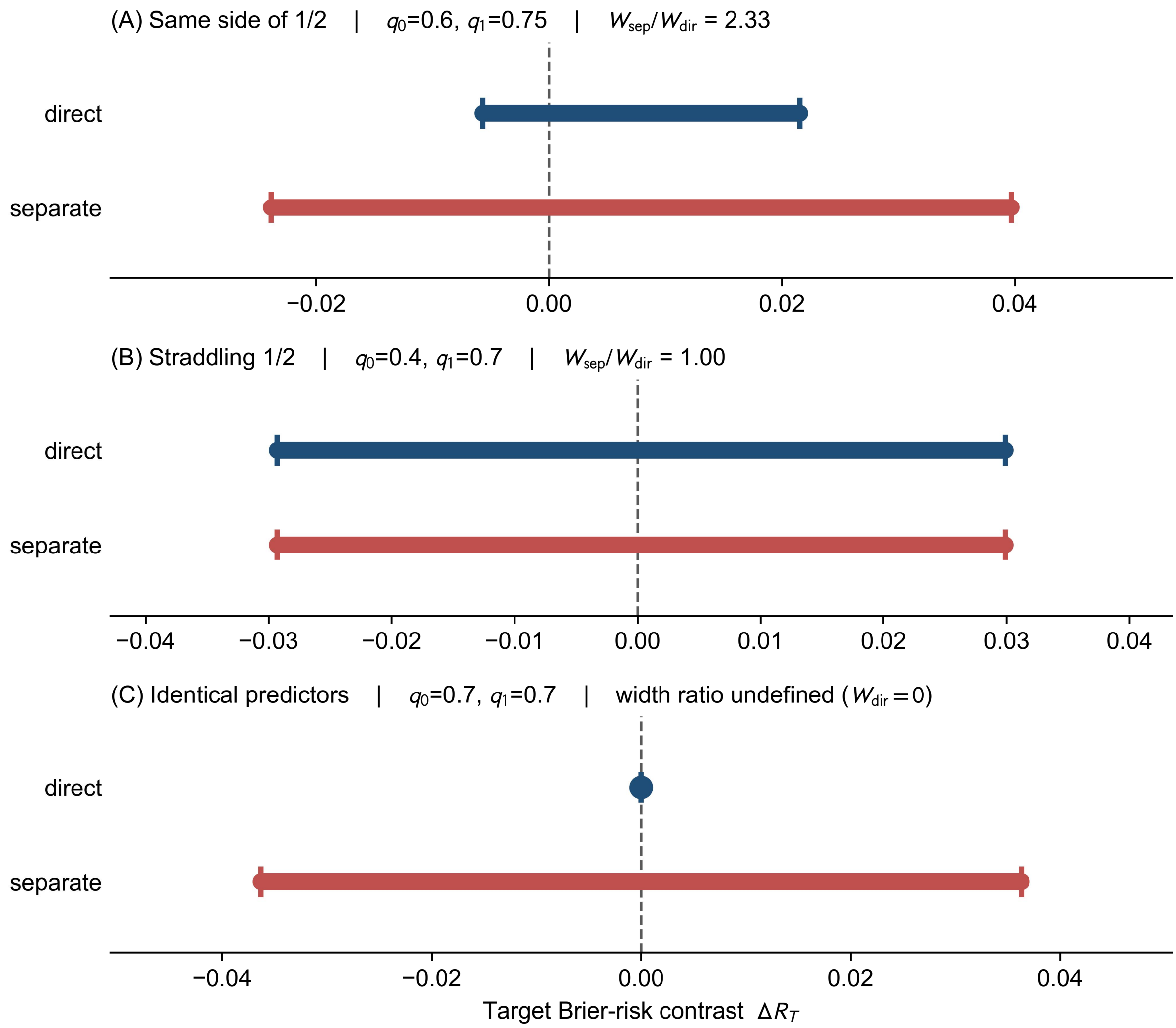

The figure shows three illustrative configurations of the two locked predictions. In each panel, the blue interval is the direct identified set $\mathcal{I}_\gamma$, which preserves the shared unidentified target outcome law, and the red interval is the separate-risk subtraction interval $\mathcal{I}_\gamma^{sep}$. The dashed vertical line marks $\Delta R_T = 0$. Panel (a) illustrates a same-side configuration $(q_0, q_1) =$ $(0.6, 0.75)$, for which $W_{sep}/W_{dir} = 2.33$ and direct identification is strictly tighter. Panel (b) illustrates a straddling configuration $(q_0, q_1) = (0.4, 0.7)$, for which $W_{sep}/W_{dir} = 1.00$. Panel (c) illustrates identical predictors $(q_0, q_1) = (0.7, 0.7)$, for which $W_{dir} = 0$, so the direct identified set collapses to the point $\Delta R_T = 0$, whereas the separate interval remains nondegenerate; the width ratio is therefore undefined. All intervals are computed from the closed-form expressions in Proposition 2.

*Alt text:* Three horizontal interval plots compare direct and separate identification, with dashed vertical lines marking zero. Panel A shows a narrower direct interval when both predictions exceed one half. Panel B shows equal widths when predictions straddle one half. Panel C shows the direct interval collapsing to zero for identical predictions, while the separate interval remains nondegenerate.

**Figure 3.** Outcome-free NSCLC evaluation comparing direct and separate identification of the target fixed-horizon Brier-risk contrast at seed 42 and $\gamma = 0.2$.

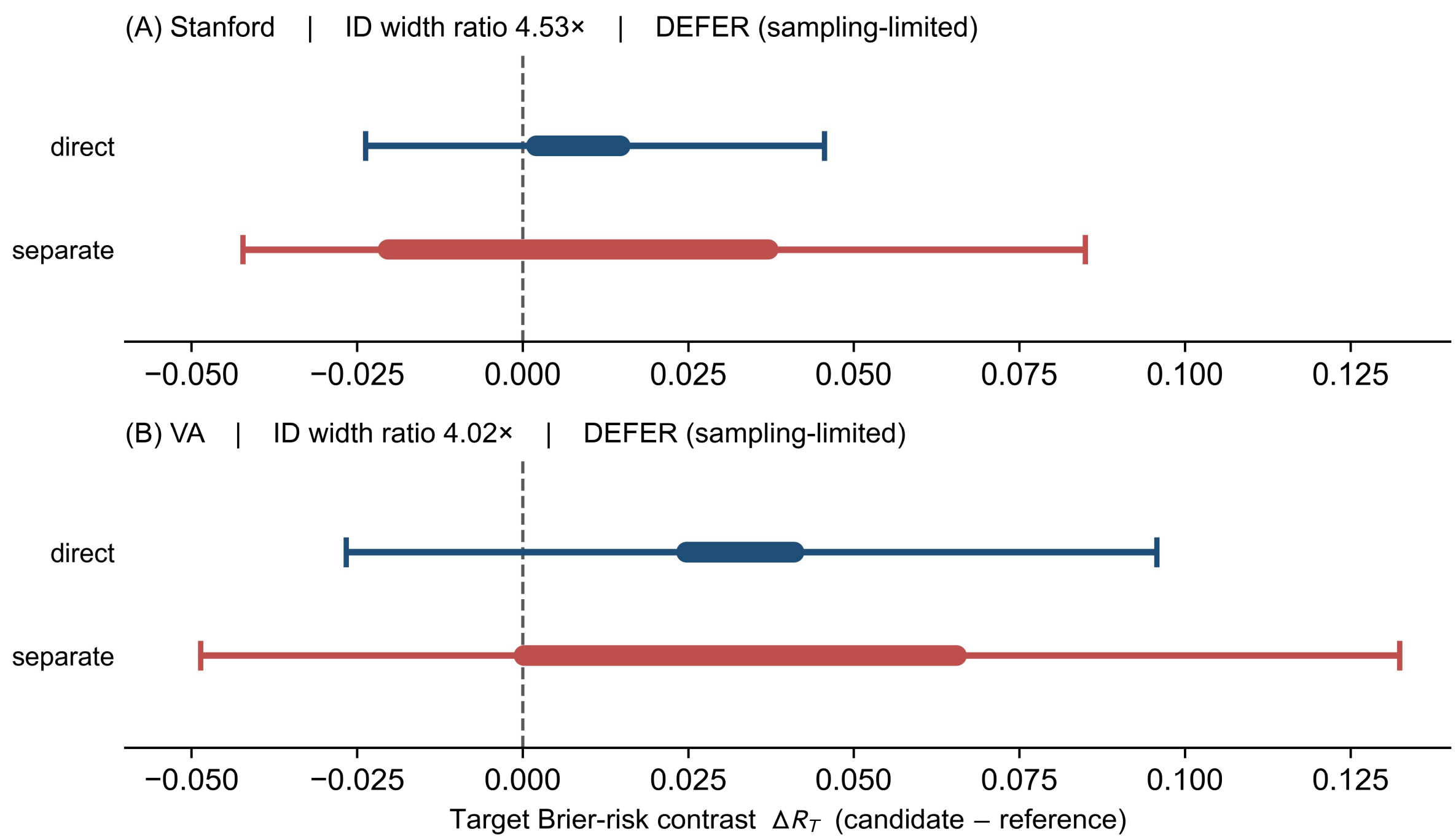

Results are shown for the Stanford and VA target cohorts. Thick bars denote the sharp identified sets, thin capped lines denote the simultaneous confidence envelopes, and the dashed vertical line marks zero. In both displayed configurations, direct identification produced substantially narrower identified sets than separate-risk subtraction, but the simultaneous confidence envelopes still included zero, leading to DEFER. The label "sampling-limited" refers to the displayed seed-42 $\gamma = 0.2$ configurations.

*Alt text:* Two panels display identified sets as thick horizontal bars and simultaneous confidence envelopes as thin capped lines for Stanford and VA. Direct identified sets are substantially narrower than separate sets in both cohorts. Although the direct identified sets lie above zero, their simultaneous confidence envelopes cross zero, resulting in DEFER for both cohorts.

**Figure 4.** Rotterdam-to-GBSG configuration with discordant decisions: direct and separate identified sets and simultaneous confidence envelopes (seed 62, $\gamma = 0.2$).

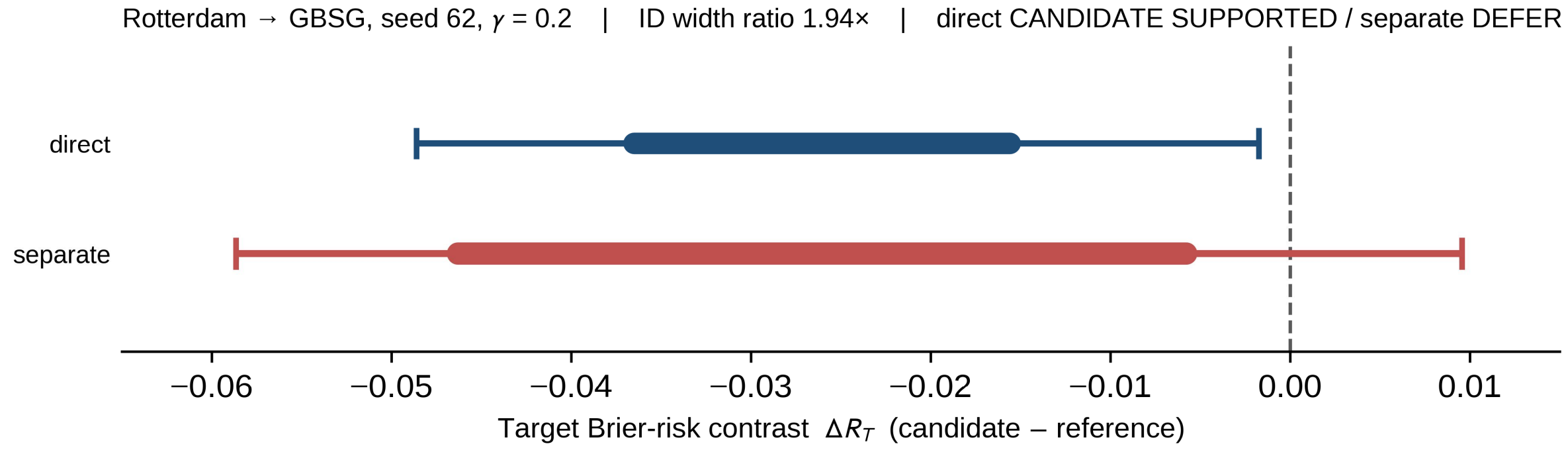

Thick segments are plug-in identified sets (direct [−0.0365, −0.0156]; separate [−0.0463, −0.0058]); thin capped lines are simultaneous confidence envelopes (direct [−0.0486, −0.0017]; separate [−0.0586, +0.0096]). The dashed vertical line marks zero. Direct yields CANDIDATE SUPPORTED; separate yields DEFER. The separate/direct identified-set width ratio is approximately 1.94.

*Alt text:* Two horizontal interval plots compare direct and separate identification for seed 62 at a sensitivity value of 0.2. Thick bars represent identified sets, thin capped lines represent simultaneous confidence envelopes, and a dashed vertical line marks zero. The direct confidence envelope lies entirely below zero, supporting candidate superiority, whereas the separate envelope crosses zero, resulting in DEFER.

**Supplementary Materials for "Sharp Partial Identification for Survival Model Comparison Without Target Outcomes" by Choi, Kim, and Kim**

## Web Appendix A. Proofs

### Web Appendix A.1. Proof of Proposition 1 (Non-identification)

Recall that

$$\Delta R_T = E_T[a(Z) - 2b(Z)m_T(Z)],$$

where

$$a(Z) = q_1^2(Z) - q_0^2(Z), \qquad b(Z) = q_1(Z) - q_0(Z).$$

Let

$$c(z) = \frac{q_0(z) + q_1(z)}{2}.$$

Then, for each $z$,

$$a(z) - 2b(z)m_T(z) = \{q_1(z) - q_0(z)\}\{q_1(z) + q_0(z) - 2m_T(z)\} = 2b(z)\{c(z) - m_T(z)\}$$

Suppose that $m_S$ is interior on the relevant target support and that

$$P_T\{q_1(Z) \neq q_0(Z)\} > 0.$$

If no finite bound is imposed on the conditional shift $\delta(z)$, then $m_T(z)$ may take any value in (0,1) at each $z$. Because the sign of $b(z)$ may vary across the target population, a uniform upward or downward perturbation of $m_T$ need not determine the sign of the population contrast. Instead, construct the target outcome law pointwise according to the sign of $b(z)$. For a target law producing a positive contrast, choose $m_T^{\mathrm{pos}}(z)$ such that

$$m_T^{\mathrm{pos}}(z) < c(z) \quad \text{when } b(z) > 0,$$

and

$$m_T^{\mathrm{pos}}(z) > c(z) \quad \text{when } b(z) < 0.$$

Then, on the set $\{z: b(z) \neq 0\}$,

$$2b(z)\{c(z) - m_T^{\mathrm{pos}}(z)\} > 0.$$

Since this set has positive $P_T$-probability,

$$\Delta R_T(m_T^{\mathrm{pos}}) > 0$$

Conversely, define $m_T^{\mathrm{neg}}(z)$ by reversing the inequalities:

$$m_T^{\mathrm{neg}}(z) > c(z) \quad \text{when } b(z) > 0,$$

and

$$m_T^{\mathrm{neg}}(z) < c(z) \quad \text{when } b(z) < 0.$$

Then

$$2b(z)\{c(z) - m_T^{\mathrm{neg}}(z)\} < 0$$

whenever $b(z) \neq 0$, and therefore

$$\Delta R_T(m_T^{\mathrm{neg}}) < 0.$$

On $\{z: b(z) = 0\}$, set

$$m_T^{\mathrm{pos}}(z) = m_T^{\mathrm{neg}}(z) = \frac{1}{2}.$$

Because $q_0(z), q_1(z) \in [0,1]$, whenever $b(z) \neq 0$ their average $c(z)$ lies strictly in (0,1). Thus $m_T(z)$ can be chosen on either side of $c(z)$ while remaining in (0,1); on $\{b(z) = 0\}$, the value of $m_T(z)$ is irrelevant to the contrast. Both constructions are therefore admissible and are observationally compatible with the observed source law and target distribution of $Z$, because target outcomes are unobserved.

Hence, in the absence of a finite restriction on the source-to-target conditional outcome shift, there exist observationally compatible target outcome laws for which $\Delta R_T > 0$ and others for which $\Delta R_T < 0$. The sign, and therefore the value, of the target Brier-risk contrast is not point identified.

In the degenerate case

$$q_0(z) = q_1(z) \qquad P_T\text{-almost surely.}$$

we have $a(z) = b(z) = 0$, and hence

$$\Delta R_T = 0$$

is trivially point identified. At the opposite extreme, when $\gamma = 0$, the sensitivity model imposes $m_T = m_S$, so the contrast is point identified under conditional transportability. This restriction, however, cannot be verified using target covariates alone.

**Remark (non-identification and sensitivity).**

Proposition 1 concerns non-identification of the sign of $\Delta R_T$ when no finite restriction is imposed on the conditional outcome shift. It should not be interpreted as a result establishing non-identification of the sensitivity parameter $\gamma$ itself. The fact that \gamma cannot be estimated or tested from the target evaluation sample follows instead from the absence of target outcomes. The survival-law realization argument in Web Appendix A.2 applies equally to these constructions.

**Web Appendix A.2. Proof of Theorem 1 (Sharp Identified Set)**

Recall that, under the sensitivity model,

$$m_-(z;\gamma) \leq m_T(z) \leq m_+(z;\gamma),$$

where

$$m_\pm(z;\gamma) = \text{expit}\{\text{logit} m_S(z) \pm \gamma\},$$

and define

$$\mathcal{M}_\gamma = \{m: m_-(z;\gamma) \leq m(z) \leq m_+(z;\gamma) \text{ for } P_T\text{-almost every } z\}.$$

From the affine representation of the target Brier-risk contrast,

$$\Delta R_T(m) = E_T[a(Z) - 2b(Z)m(Z)],$$

with

$$a(Z) = q_1^2(Z) - q_0^2(Z), \;\; b(Z) = q_1(Z) - q_0(Z).$$

For each fixed $z$, the integrand

$$a(z) - 2b(z)m(z)$$

is affine in $m(z)$. Because the admissible values of $m(z)$ form the pointwise interval

$$[m_-(z;\gamma), m_+(z;\gamma)],$$

the pointwise maximum is obtained at the endpoint determined by the sign of $-2b(z)$. Thus, define

$$m_U^*(z) = \begin{cases} m_-(z;\gamma), & b(z) \geq 0, \\ m_+(z;\gamma), & b(z) < 0, \end{cases}$$

and

$$m_L^*(z) = \begin{cases} m_+(z;\gamma), & b(z) \geq 0, \\ m_-(z;\gamma), & b(z) < 0, \end{cases}$$

it follows that

$$U_\gamma = E_T[a(Z) - 2b(Z)m_U^*(Z)] = \sup_{m \in \mathcal{M}_\gamma} E_T[a(Z) - 2b(Z)m(Z)],$$

and

$$L_\gamma = E_T[a(Z) - 2b(Z)m_L^*(Z)] = \inf_{m \in \mathcal{M}_\gamma} E_T[a(Z) - 2b(Z)m(Z)].$$

Because $b$, $m_-$, and $m_+$ are measurable, both $m_U^*$ and $m_L^*$ are measurable. Both extrema are attainable. Indeed, $m_U^*$ and $m_L^*$ select only the admissible pointwise endpoints $m_-$ and $m_+$, corresponding to conditional log-odds shifts equal to $-\gamma$ or $+\gamma$. Hence both functions belong to $\mathcal{M}_\gamma$.

It remains to establish that every value between $L_\gamma$ and $U_\gamma$ is also attainable. For any $t \in [0,1]$, define

$$m_t(z) = (1-t)m_L^*(z) + t m_U^*(z).$$

Because the constraint is pointwise rectangular,

$$m_t(z) \in [m_-(z;\gamma), m_+(z;\gamma)]$$

for every $t \in [0,1]$, so $m_t \in \mathcal{M}_\gamma$. Moreover, because the logit function is strictly increasing,

$$\delta_t(z) = \mathrm{logit} m_t(z) - \mathrm{logit} m_S(z)$$

satisfies $|\delta_t(z)| \le \gamma$. Thus $m_t$ satisfies the original sensitivity model, not merely the probability-scale interval constraint.

By affinity,

$$\Delta R_T(m_t) = (1-t)\Delta R_T(m_L^*) + t\Delta R_T(m_U^*) = (1-t)L_\gamma + tU_\gamma.$$

Therefore every value in

$$[L_\gamma,\ U_\gamma]$$

is generated by an admissible target conditional outcome law. Consequently,

$$\mathcal{I}_\gamma = [L_\gamma,\ U_\gamma]$$

is the sharp identified set.

Write $d_\gamma(z) = m_+(z;\gamma) - m_-(z;\gamma)$. Subtracting the two endpoint expressions gives

$$U_\gamma - L_\gamma = E_T[2|b(Z)|\{m_+(Z;\gamma) - m_-(Z;\gamma)\}],$$

and therefore

$$\boxed{W_{\mathrm{dir}} = 2E_T\left[|q_1(Z) - q_0(Z)|\, d_\gamma(Z)\right]}.$$

**Survival-law Realization of Admissible Fixed-horizon Probabilities**

The sharpness result above is not merely a statement about an abstract binary outcome law at the horizon $\tau$. Every admissible conditional survival probability $m(z) \in (0,1)$ can be realized by a proper event-time distribution.

For any $m \in \mathcal{M}_\gamma$, define

$$\lambda(z) = -\frac{\log m(z)}{\tau} > 0$$

and let

$$T^* \mid Z = z \sim \text{Exponential}\{\lambda(z)\}.$$

Then

$$P(T^* > \tau \mid Z = z) = \exp\{-\lambda(z)\tau\} = \exp\{\log m(z)\} = m(z).$$

Hence every admissible $m(z)$ corresponds to a valid conditional survival-time law. A censoring distribution satisfying the conditional independent-censoring assumption may then be attached separately. Thus the identified set is sharp at the level of full survival-time laws, not only at the level of the fixed-horizon binary outcome. This exponential construction is used solely to establish existence for sharpness and is not an assumed event-time model for estimation.

**Web Appendix A.3. Proof of Proposition 2 (Direct Versus Separate Identification)**

For $j \in \{0,1\}$, write the target Brier risk as

$$R_T(q_j) = E_T[q_j^2(Z) + \{1 - 2q_j(Z)\}m_T(Z)].$$

Let

$$[L_{j,\gamma}, U_{j,\gamma}]$$

denote the sharp identified set for $R_T(q_j)$ under the same sensitivity class $\mathcal{M}_\gamma$, and define the separate-risk subtraction interval as

$$\mathcal{I}_\gamma^{\text{sep}} = [L_{1,\gamma} - U_{0,\gamma},\ U_{1,\gamma} - L_{0,\gamma}].$$

The direct identified set is

$$\mathcal{I}_\gamma = [L_\gamma,\ U_\gamma].$$

**(i) Set inclusion**

Under direct identification, both component risks are evaluated under the same admissible target outcome law $m_T$. Thus the feasible pairs of component risks are restricted to

$$\{(R_T(q_1; m),\ R_T(q_0; m)) : m \in \mathcal{M}_\gamma\}.$$

By contrast, separate-risk subtraction allows the two component risks to attain their extrema under potentially different admissible laws,

$$\{(R_T(q_1; m_1),\ R_T(q_0; m_0)) : m_1, m_0 \in \mathcal{M}_\gamma\}.$$

Define

$$\mathcal{F}_{\text{dir}} = \{(R_T(q_1; m), R_T(q_0; m)): m \in \mathcal{M}_\gamma\}$$

and

$$\mathcal{F}_{\text{sep}} = \{(R_T(q_1; m_1), R_T(q_0; m_0)): m_1, m_0 \in \mathcal{M}_\gamma\}$$

Every pair in $\mathcal{F}_{\text{dir}}$ is obtained in $\mathcal{F}_{\text{sep}}$ by setting $m_1 = m_0 = m$. Hence

$$\mathcal{F}_{\text{dir}} \subseteq \mathcal{F}_{\text{sep}}$$

Applying the contrast map $D(r_1, r_0) = r_1 - r_0$ therefore yields

$$\boxed{\mathcal{I}_\gamma \subseteq \mathcal{I}_\gamma^{\text{sep}}.}$$

The inclusion reflects preservation of the shared unidentified target outcome law in the direct procedure.

**(ii) Closed-form widths**

Define $d_\gamma(z) = m_+(z; \gamma) - m_-(z; \gamma)$. For each component risk, the coefficient multiplying $m_T(Z)$ is

$$1 - 2q_j(Z).$$

By the same pointwise extremization argument as in Theorem 1,

$$U_{j,\gamma} - L_{j,\gamma} = E_T\big[|1 - 2q_j(Z)|\, d_\gamma(Z)\big].$$

Hence the width of the separate-risk subtraction interval is

$$\boxed{W_{\text{sep}} = E_T\big[(|1 - 2q_1(Z)| + |1 - 2q_0(Z)|)\, d_\gamma(Z)\big]}.$$

From Theorem 1, the width of the direct identified set is

$$\boxed{W_{\text{dir}} = 2E_T\big[|q_1(Z) - q_0(Z)|\, d_\gamma(Z)\big]}.$$

**(iii) Width ordering and strictness**

Pointwise,

$$|1 - 2q_1| + |1 - 2q_0| \geq |(1 - 2q_1) - (1 - 2q_0)| = 2|q_1 - q_0|,$$

by the triangle inequality. Multiplying by the nonnegative factor

$$d_\gamma(Z)$$

and taking expectation yields

$$\boxed{W_{\mathrm{dir}} \le W_{\mathrm{sep}}}.$$

Equality in the pointwise triangle inequality holds exactly when

$$\big(1 - 2q_1(Z)\big)\big(1 - 2q_0(Z)\big) \le 0,$$

including the boundary case where either prediction equals 1/2. Equivalently, $q_0(Z)$ and $q_1(Z)$ lie on opposite sides of 1/2, or at least one equals 1/2.

Therefore, on the region where the sensitivity band is nondegenerate,

$$\{d_\gamma(Z) > 0\}$$

global equality of the two widths holds if and only if

$$\big(1 - 2q_1(Z)\big)\big(1 - 2q_0(Z)\big) \le 0$$

for $P_T$-almost every $Z$ in that region.

Equivalently, strict tightening occurs if and only if

$$\boxed{P_T\big(\{(q_1(Z) - 1/2)(q_0(Z) - 1/2) > 0\} \cap \{d_\gamma(Z) > 0\}\big) > 0}.$$

$$W_{\mathrm{sep}} - W_{\mathrm{dir}} = 2E_T\big[\min\{|1 - 2q_0(Z)|, |1 - 2q_1(Z)|\}\, \mathbf{1}\{(q_0(Z) - 1/2)(q_1(Z) - 1/2) > 0\}\, d_\gamma(Z)\big].$$

Thus, the excess width induced by separate-risk subtraction is exactly the contribution from target profiles on which both predictions lie on the same side of 1/2.

When $\gamma = 0$,

$$m_+(Z;\gamma) = m_-(Z;\gamma) = m_S(Z),$$

so both widths equal zero regardless of the geometry of $q_0$ and $q_1$.

**Illustrative Special Case**

Suppose

$$q_0(Z) = q_1(Z) \qquad P_T\text{-almost surely.}$$

Then the Brier-risk contrast is identically zero,

$$\Delta R_T = 0,$$

and hence

$$W_{\text{dir}} = 0$$

However, if

$$P_T\{q_0(Z) \neq 1/2\,,\ d_\gamma(Z) > 0\} > 0,$$

then

$$W_{\text{sep}} = E_T\big[2|1 - 2q_0(Z)|d_\gamma(Z)\big] > 0$$

Thus, separately identifying the two component risks can introduce identification uncertainty that is not intrinsic to the contrast itself.

**Web Appendix A.4. Proof of Proposition 3 (Robustness Value)**

For each $\gamma \geq 0$, let

$$\mathcal{M}_\gamma = \{m\colon m_-(z;\gamma) \leq m(z) \leq m_+(z;\gamma)\ \text{ for } P_T\text{-almost every } z\},$$

where

$$m_\pm(z;\gamma) = \text{expit}\{\text{logit} m_S(z) \pm \gamma\}.$$

If $0 \leq \gamma_1 \leq \gamma_2$, then for every $z$,

$$m_-(z;\gamma_2) \leq m_-(z;\gamma_1) \leq m_+(z;\gamma_1) \leq m_+(z;\gamma_2).$$

Hence

$$\mathcal{M}_{\gamma_1} \subseteq \mathcal{M}_{\gamma_2}.$$

The sharp lower and upper endpoints are

$$L_\gamma = \inf_{m\in\mathcal{M}_\gamma} \Delta R_T(m), \qquad U_\gamma = \sup_{m\in\mathcal{M}_\gamma} \Delta R_T(m).$$

Because the admissible class expands with $\gamma$,

$$L_{\gamma_2} \leq L_{\gamma_1},$$

and

$$U_{\gamma_2} \geq U_{\gamma_1}.$$

Therefore

$$\boxed{L_\gamma \text{ is nonincreasing in } \gamma, \qquad U_\gamma \text{ is nondecreasing in } \gamma}.$$

This establishes the nesting of the sharp identified sets,

$$[L_{\gamma_1},\ U_{\gamma_1}] \subseteq [L_{\gamma_2},\ U_{\gamma_2}] \qquad \text{whenever } \gamma_1 \leq \gamma_2.$$

For each $z$, $m_-(z;\gamma)$ and $m_+(z;\gamma)$ are continuous in $\gamma$. Because the endpoint integrands are dominated by an integrable envelope, the dominated convergence theorem implies that

$$\boxed{L_\gamma \text{ and } U_\gamma \text{ are continuous in } \gamma}.$$

**Candidate Superiority**

Candidate superiority is identified at sensitivity value $\gamma$ whenever

$$U_\gamma < 0,$$

because every target law admitted by $\mathcal{M}_\gamma$ then yields

$$\Delta R_T < 0.$$

Define

$$\mathcal{C}_{\text{cand}} = \{\gamma \geq 0 : U_\gamma < 0\}.$$

Since $U_\gamma$ is nondecreasing, if $\gamma_2 \in \mathcal{C}_{\text{cand}}$ and $0 \leq \gamma_1 \leq \gamma_2$, then

$$U_{\gamma_1} \leq U_{\gamma_2} < 0.$$

Thus,

$$\gamma_1 \in \mathcal{C}_{\text{cand}}.$$

Hence $\mathcal{C}_{\text{cand}}$ is downward-closed. The population robustness value for candidate superiority may therefore be defined as

$$\boxed{\gamma^*_{\text{cand}} = \sup\{\gamma \geq 0 : U_\gamma < 0\}}$$

The boundary need not be characterized by a unique sign-changing root because $U_\gamma$ may equal zero on an interval. By continuity, if $\mathcal{C}_{\text{cand}} \neq \varnothing$ and $\gamma^*_{\text{cand}} < \infty$, then

$$U_{\gamma^*_{\text{cand}}} = 0.$$

Because certification is defined by the strict inequality $U_\gamma < 0$, the supremum $\gamma^*_{\text{cand}}$ need not itself belong to $\mathcal{C}_{\text{cand}}$. A finite prespecified $\gamma$-grid need not contain this population boundary exactly.

**Reference Superiority**

Similarly, reference superiority is identified whenever

$$L_\gamma > 0.$$

Define

$$\mathcal{C}_{\text{ref}} = \{\gamma \geq 0 : L_\gamma > 0\}.$$

Because $L_\gamma$ is nonincreasing, if $\gamma_2 \in \mathcal{C}_{\text{ref}}$ and $0 \leq \gamma_1 \leq \gamma_2$, then

$$L_{\gamma_1} \geq L_{\gamma_2} > 0,$$

so $\gamma_1 \in \mathcal{C}_{\text{ref}}$. Hence $\mathcal{C}_{\text{ref}}$ is also downward-closed, and the corresponding robustness value is

$$\boxed{\gamma^*_{\text{ref}} = \sup\{\gamma \geq 0 : L_\gamma > 0\}}$$

Similarly, if $\mathcal{C}_{\text{ref}} \neq \varnothing$ and $\gamma^*_{\text{ref}} < \infty$, continuity implies

$$L_{\gamma^*_{\text{ref}}} = 0$$

Because certification is defined by the strict inequality $L_\gamma > 0$, the supremum $\gamma^*_{\text{ref}}$ need not itself belong to $\mathcal{C}_{\text{ref}}$.

As in Section 2.7, an empty certification set is assigned the value zero, and persistence of the sign to the upper end of the examined range is reported as exceeding that range.

Thus, the robustness value characterizes the boundary of the sensitivity range over which the sign of the sharp population identified set remains fixed. It is an identification-level sensitivity summary and is not an estimator of the realized source-to-target conditional shift.

**Remark on Sampling-aware Certification**

The proposition concerns the population identified-set endpoints $L_\gamma$ and $U_\gamma$. In finite samples, directional certification is instead based on the simultaneous confidence envelope. For example, candidate superiority is certified only when the simultaneous upper envelope lies below zero. The corresponding sampling-aware certified range is therefore generally no larger than the population robustness range on the simultaneous-coverage event.

On the simultaneous-coverage event,

$$\{\gamma: \overline{U}_\gamma < 0\} \subseteq \{\gamma: U_\gamma < 0\}, \ \{\gamma: \underline{L}_\gamma > 0\} \subseteq \{\gamma: L_\gamma > 0\}.$$

**Web Appendix A.5. Proof of Theorem 2 (Simultaneous Inference for the Identified-set Endpoints)**

Let $n_S$ and $n_T$ denote the source and target sample sizes, respectively, let

$$n = n_S + n_T, \qquad \frac{n_T}{n} \to \rho \in (0,1),$$

and let

$$\mathcal{G} = \{\gamma_1, \ldots, \gamma_K\}, \qquad K < \infty,$$

denote the prespecified sensitivity grid. Define the stacked endpoint vector

$$\vartheta = \left(L_{\gamma_1}, U_{\gamma_1}, \cdots L_{\gamma_K}, U_{\gamma_K}\right)^\top \in \mathbb{R}^{2K},$$

with corresponding estimator $\hat{\vartheta}$.

Conditions (A1)–(A4) of Theorem 2 in the main text correspond to (R1); (R2), (R3), and (R10); (R4)–(R6) and (R9); and (R8) and (R11), respectively, with the finite grid (R7) part of the statement.

**Regularity Conditions**

The result is established under the following conditions.

(R1) Locked prediction functions.

The prediction functions $q_0$ and $q_1$ are fixed measurable functions and are not re-estimated from either the source calibration sample or the target evaluation sample.

(R2) Correct source outcome model and source standardization.

Let $\xi_0$ denote the finite-dimensional source-calibration standardization parameters used to construct the standardized deployment-summary variables, and write

$$m_S(z;\theta_0,\xi_0) = \text{expit}\{\theta_0^\top \phi(z;\xi_0)\}.$$

Let $\nu = (\theta, \xi)$, where $\phi(z;\xi)$ is the source-standardized logistic design vector and $\xi$ contains its standardization parameters. The IPCW-weighted logistic estimator and the empirical source-standardization estimator are assumed to satisfy

$$\sqrt{n_S}(\hat{\nu} - \nu_0) = \frac{1}{\sqrt{n_S}} \sum_{i=1}^{n_S} \psi_\nu\left(O_{S,i}\right) + o_p(1).$$

with

$$E_S\{\psi_\nu(O_S)\} = 0,\ \ E_S\|\psi_\nu(O_S)\|^2 < \infty.$$

The influence function $\psi_\nu$ includes the first-order effect of estimating the censoring nuisance quantities used in the IPCW weights.

(R3) Conditional independent censoring and correct censoring model

Let $H = (Z, V_C)$ denote the prespecified covariate vector used in the source censoring model, where $V_C$ may contain additional censoring-related covariates. In the primary analyses, $V_C$ is empty and hence $H = Z$. Assume

$$T^* \perp C|H.$$

The source censoring distribution is modeled through

$$\lambda_C(t|H) = \lambda_{C0}(t)\exp(\alpha_0^\top H),$$

with $\hat{\alpha}$ estimated by Cox partial likelihood and $\hat{\Lambda}_{C0}$ by the Breslow estimator (Breslow, 1972; Cox, 1972). The Cox–Breslow estimator is uniformly consistent on $[0,\tau]$ and admits its standard martingale asymptotic representation. The joint estimation of $\left(\nu, \alpha, \hat{\Lambda}_{C0}\right)$ is assumed regular (Andersen and Gill, 1982).

(R4) Censoring positivity.

$$\inf_{t \le \tau, h} G_0(t \mid h) \ge \epsilon_G > 0.$$

(R5) Interiority.

For some $\epsilon_m > 0$,

$$\epsilon_m \le m_S(z) \le 1 - \epsilon_m$$

on the relevant support.

(R6) Transport positivity and moments.

The target distribution of $Z$ is absolutely continuous with respect to its source distribution and has an essentially bounded density ratio,

$$P_T^Z \ll P_S^Z, \ \frac{dP_T^Z}{dP_S^Z}(z) \le B_0 < \infty.$$

In addition,

$$E_T\{a(Z)^2\} < \infty, \ \ E_T\{b(Z)^2\} < \infty.$$

(R7) Finite prespecified sensitivity grid.

The set $\mathcal{G}$ is fixed and finite, with $K < \infty$.

(R8) Regularity of the IPCW endpoint functional.

The fixed-horizon IPCW estimating map and the maps from the nuisance parameters to the endpoint functionals are differentiable in the required functional sense, with finite second moments under (R4)–(R6).

(R9) Asymptotically inactive numerical truncation.

If estimated censoring probabilities are truncated below at 0.02 in the numerical implementation, assume

$$\inf_{t \le \tau, h} G_0(t \mid h) > 0.02$$

and

$$\sup_{t\le\tau,\,h}\left|\hat{G}(t|h)-G_0(t|h)\right|=o_p(1).$$

The positive margin above 0.02 and uniform consistency therefore imply that the numerical truncation is inactive with probability tending to one.

(R10) Sampling and empirical-process conditions.

The source sample and target sample are independent, with observations independently and identically distributed within each population. The function classes required for the nuisance estimators and endpoint expansions satisfy the relevant Donsker and stochastic-equicontinuity conditions.

(R11) Positive marginal variances.

Every distinct endpoint coordinate retained in the studentized maximum has strictly positive asymptotic marginal variance. At $\gamma=0$, where $L_0=U_0$, the duplicated coordinate may be represented by a single endpoint coordinate with positive marginal variance.

**Step 1. Asymptotic linearity of each endpoint**

We write $P_T f=E_T f$ and $P_{n_T}$ for the target empirical measure.

Consider first the upper endpoint,

$$U_\gamma=P_T[a(Z)-2b(Z)m_U^*(Z)],$$

with estimator

$$\hat{U}_\gamma=P_{n_T}[a(Z)-2b(Z)\hat{m}_U^*(Z)].$$

Because $q_0$ and $q_1$ are locked under (R1),

$$b(z)=q_1(z)-q_0(z)$$

is fixed. Hence the pointwise corner assignment defining $m_U^*(Z)$ is determined by the fixed sign of $b(z)$ and does not introduce an estimated nonsmooth optimization boundary.

In particular, on

$$\{z: b(z)=0\},$$

the coefficient multiplying the unknown outcome probability is

$$-2b(z) = 0.$$

The endpoint functional is therefore invariant to which admissible corner is assigned on this set. Consequently, no condition such as

$$P_T\{b(Z) = 0\} = 0$$

is required.

For either sensitivity boundary,

$$m_\pm(z;\gamma) = \text{expit}\{\text{logit} m_S(z) \pm \gamma\},$$

the derivative with respect to $m_S$ is

$$\dot{m}_\pm(z) = \frac{\text{expit}'\{\text{logit} m_S(z) \pm \gamma\}}{m_S(z)\{1 - m_S(z)\}}.$$

By (R5), this derivative is uniformly bounded over the relevant support and over the finite grid $\mathcal{G}$. The functional delta method therefore gives

$$\|\widehat{m}_\pm - m_\pm - \dot{m}_\pm(\widehat{m}_S - m_S)\|_{L_2(P_T)} = o_p\left({n_S}^{-1/2}\right).$$

Define

$$h_{U,\gamma}(z;\nu) = a(z) - 2b(z)m_U^*(z;\nu)$$

Then

$$\widehat{U}_\gamma - U_\gamma = (P_{nT} - P_T)h_{U,\gamma}(\bullet;\nu_0) + P_T\{h_{U,\gamma}(\bullet;\hat{\nu}) - h_{U,\gamma}(\bullet;\nu_0)\}$$

$$+ (P_{nT} - P_T)\{h_{U,\gamma}(\bullet;\hat{\nu}) - h_{U,\gamma}(\bullet;\nu_0)\}.$$

Under (R10) and the independence of the source and target samples, the final empirical-process term is $o_p(n^{-1/2})$.

Combining the preceding expansion with the source outcome and standardization nuisance representation in (R2) yields

$$\sqrt{n}\left(\widehat{U}_\gamma - U_\gamma\right) = \sqrt{n}P_{n_T}\left[\text{IF}_{U,\gamma}^T\right] + \sqrt{n}P_{n_S}\left[\text{IF}_{U,\gamma}^S\right] + o_p(1),$$

where the centered target contribution is

$$\mathrm{IF}^{T}_{U,\gamma}(Z) = a(Z) - 2b(Z)m^{*}_{U}(Z) - U_{\gamma}.$$

To express the source contribution, define

$$\Gamma_{U,\gamma} = -2E_{T}[b(Z)\dot{m}^{*}_{U}(Z)\nabla_{\nu}m_{S}(Z)].$$

where $\dot{m}^{*}_{U}(Z)$ denotes the derivative with respect to $m_S$ of the sensitivity-boundary probability selected by the upper-endpoint corner rule, and

$$\nabla_{\nu}m_{S}(Z)$$

is the gradient of the fitted source conditional survival probability with respect to

$$\nu = (\theta, \xi).$$

The source influence-function contribution is therefore

$$\boxed{\mathrm{IF}^{S}_{U,\gamma}(O_S) = \Gamma^{\top}_{U,\gamma}\psi_{\nu}(O_S)}.$$

By construction, $\psi_{\nu}$ already incorporates the first-order effect of estimating the conditional censoring model used to form the IPCW weights, so no additional censoring-nuisance contribution is added separately.

The same argument applies to $L_{\gamma}$, with the corresponding lower-endpoint corner rule. Thus, each endpoint estimator is asymptotically linear with one target empirical contribution and one source nuisance-estimation contribution.

The differentiability step above is an application of the functional delta method (Van Der Vaart and Wellner, 1996).

**Step 2. Joint Asymptotic Normality Over the Finite Grid**

Because $K < \infty$, stacking the expansions from Step 1 gives a finite-dimensional asymptotic linear representation for

$$\sqrt{n}(\hat{\vartheta} - \vartheta).$$

The source and target samples are independent. Consequently, the limiting covariance matrix has the additive form

$$\Sigma = \rho^{-1}\Sigma_T + (1-\rho)^{-1}\Sigma_S.$$

where $\Sigma_T$ and $\Sigma_S$ denote the covariance matrices of the stacked target and source influence-function contributions, respectively.

Under the moment and empirical-process conditions in (R6) and (R10), the multivariate central limit theorem gives

$$\boxed{\sqrt{n}\left(\hat{\vartheta} - \vartheta\right) \overset{d}{\to} N(0, \Sigma).}$$

No high-dimensional Gaussian approximation is required because the dimension $2K$ is fixed.

**Step 3. Consistency of the Joint Pairs Bootstrap**

The bootstrap independently resamples the source and target observations with replacement. Within each source bootstrap sample, all estimated nuisance quantities are refitted, including the conditional Cox censoring regression, the Breslow baseline cumulative hazard, the source-score standardization parameters, and the IPCW-weighted logistic outcome regression.

The locked prediction functions $q_0$ and $q_1$ remain fixed.

Under (R2), (R3), and (R10), the source estimators constitute regular semiparametric M- or Z-estimators. Standard exchangeably weighted bootstrap results for semiparametric estimators therefore imply conditional weak convergence of their bootstrap versions (Cheng and Huang, 2010). Passing from the nuisance estimators to the endpoint functionals additionally uses the bootstrap functional delta method (Van Der Vaart and Wellner, 1996) Consequently, conditionally on the observed data,

$$\boxed{\sqrt{n}\left(\hat{\vartheta}^* - \hat{\vartheta}\right) \overset{d^*}{\to} N(0, \Sigma)}$$

in probability.

It follows that the bootstrap standard deviations satisfy

$$\hat{s}_{L,\gamma} = \mathrm{sd}^*\left(\hat{L}_\gamma^*\right) = \frac{\sigma_{L,\gamma}}{\sqrt{n}} + o_p\left(n^{-1/2}\right),$$

and similarly,

$$\hat{s}_{U,\gamma} = \frac{\sigma_{U,\gamma}}{\sqrt{n}} + o_p\left(n^{-1/2}\right).$$

Thus, no additional $1/\sqrt{n}$ factor is applied to the bootstrap standard errors.

**Step 4. Studentized Maximum and Simultaneous Envelope**

In bootstrap replicate $r$, define the studentized maximum

$$\widehat{M}^{*r} = \max_{\gamma \in \mathcal{G}} \max\left\{\frac{\left|\hat{L}_\gamma^{*r} - \hat{L}_\gamma\right|}{\hat{s}_{L,\gamma}}, \frac{\left|\widehat{U}_\gamma^{*r} - \widehat{U}_\gamma\right|}{\hat{s}_{U,\gamma}}\right\}.$$

Because $\mathcal{G}$ is finite, Steps 2 and 3, together with consistency of the bootstrap standard errors, allow direct application of the continuous mapping theorem and Slutsky's theorem. After removal of any duplicated $\gamma = 0$ coordinate,

$$\widehat{M}^{*r} \xrightarrow{d^*} M_\infty = \max_j \left|\zeta_j\right|,$$

where

$$\zeta \sim N(0, R),$$

and $R$ is the correlation matrix of the distinct standardized endpoint coordinates.

Under (R11), the limiting maximum has a continuous distribution at its relevant quantile. Let $c_{1-\alpha}$ denote the $(1-\alpha)$-quantile of the limiting distribution, and let $c_{1-\alpha}^*$ denote the ideal conditional $(1-\alpha)$-quantile of the bootstrap statistic $\widehat{M}^{*r}$. Bootstrap consistency implies

$$c_{1-\alpha}^* \xrightarrow{p} c_{1-\alpha}.$$

Let $\hat{c}_{1-\alpha,B}$ denote the empirical quantile computed from $B$ bootstrap replications. When $B \to \infty$,

$$\hat{c}_{1-\alpha,B} - c_{1-\alpha}^* \xrightarrow{p} 0.$$

Write $\hat{c}_{1-\alpha} \equiv \hat{c}_{1-\alpha,B}$ for the quantile computed from $B$ bootstrap replicates. When $B \to \infty$, it is asymptotically equivalent to the ideal conditional bootstrap quantile.

Define

$$\underline{L}_\gamma = \hat{L}_\gamma - \hat{c}_{1-\alpha}\hat{s}_{L,\gamma},$$

and

$$\overline{U}_\gamma = \hat{U}_\gamma + \hat{c}_{1-\alpha}\hat{s}_{U,\gamma}.$$

Let $\widehat{M}$ denote the corresponding studentized maximum formed from the sampling errors of the original endpoint estimators. The event

$$\{\widehat{M} \le \hat{c}_{1-\alpha}\}$$

implies simultaneously that

$$L_\gamma \ge \underline{L}_\gamma$$

and

$$U_\gamma \le \overline{U}_\gamma$$

for every $\gamma \in \mathcal{G}$. Therefore,

$$P\{L_\gamma \ge \underline{L}_\gamma,\ U_\gamma \le \overline{U}_\gamma \text{ for all } \gamma \in \mathcal{G}\}$$

$$\ge P\{\widehat{M} \le \hat{c}_{1-\alpha}\}$$

$$= 1 - \alpha + o(1).$$

Hence,

$$\boxed{P\{L_\gamma \ge \underline{L}_\gamma,\ U_\gamma \le \overline{U}_\gamma \text{ for all } \gamma \in \mathcal{G}\} \ge 1 - \alpha + o(1).}$$

The inequality rather than equality is intentional. The desired outer-envelope event is one-sided for each endpoint: only downward coverage of $L_\gamma$ and upward coverage of $U_\gamma$ are required. By contrast, $\widehat{M}$ is constructed using two-sided absolute studentized deviations. The resulting outer envelope can therefore be conservative.

**Step 5. The Duplicated Coordinate at $\boldsymbol{\gamma = 0}$**

At

$$\gamma = 0,$$

we have

$$m_-(z;0) = m_+(z;0) = m_S(z)$$

and consequently

$$L_0 = U_0$$

The two endpoint estimators at $\gamma = 0$ are likewise identical, so the covariance matrix of the full stacked vector is necessarily singular.

This singularity does not invalidate the argument.

The two $\gamma = 0$ studentized terms are exact duplicates. Removing either duplicate leaves the maximum statistic unchanged. Thus, validity does not require the full covariance matrix $\Sigma$ to be nonsingular. It is sufficient that the distinct retained endpoint coordinates have positive marginal variances as required by (R11).

After removal of the duplicate coordinate, the finite-dimensional CLT, bootstrap convergence, continuous mapping argument, and quantile consistency in Step 4 remain unchanged.

This completes the proof of Theorem 2.

**Remark (continuous sensitivity ranges).**

Extension to a continuous sensitivity range is possible under additional uniform empirical-process and bootstrap regularity conditions, but the present theory and implementation focus on a finite prespecified sensitivity grid.

**Web Appendix A.6. First-Order limit of the scalar-to-profile-varying width ratio**

For $\eta \in \mathbb{R}$, let $m_\eta(z) = \text{expit}\{\text{logit} m_S(z) + \eta\}$ denote the $Z$-conditional constant tilt, and let $\Delta R(\eta) = E_T\{a(Z) - 2b(Z)m_\eta(Z)\}$. For $\gamma \geq 0$, define the scalar range width $W_{\text{scal}}(\gamma) =$

$\max_{|\eta|\le\gamma}\Delta R(\eta) - \min_{|\eta|\le\gamma}\Delta R(\eta)$. Let $w(z) = m_S(z)\{1 - m_S(z)\}$ and assume $E_T\{|b(Z)|w(Z)\} > 0$. Profiles with $m_S(z) \in \{0,1\}$ have $w(z) = 0$ and $d_\gamma(z) = 0$ and contribute to neither width.

**Claim.** As $\gamma \downarrow 0$,

$$\frac{W_{\mathrm{scal}}(\gamma)}{W_{\mathrm{dir}}(\gamma)} \to \kappa = \frac{|E_T\{b(Z)w(Z)\}|}{E_T\{|b(Z)|w(Z)\}}.$$

*Proof.* Write $g = \mathrm{expit}$ and $\ell(z) = \mathrm{logit} m_S(z)$, so that $g' = g(1-g)$, $g'' = g'(1-2g)$, and $g''' = g'(1-6g')$. Since $0 \le g' \le 1/4$, we have $\sup|g''| = 1/(6\sqrt{3}) < 1/10$ and $\sup|g'''| = 1/8$.

**Step 1 (direct width).** By Proposition 2, $W_{\mathrm{dir}}(\gamma) = 2E_T\{|b(Z)|d_\gamma(Z)\}$ with $d_\gamma = g(\ell+\gamma) - g(\ell-\gamma)$. A third-order Taylor expansion of $g(\ell \pm \gamma)$ about $\ell$ gives $|d_\gamma(z) - 2\gamma w(z)| \le \gamma^3/24$ uniformly in $z$. Since $|b| \le 1$, $W_{\mathrm{dir}}(\gamma) = 4\gamma E_T(|b|w) + r_1(\gamma)$ with $|r_1(\gamma)| \le \gamma^3/12$.

**Step 2 (scalar width).** Since $\partial m_\eta/\partial\eta = g'(\ell+\eta)$ and $|b| \le 1$, dominated convergence gives $\Delta R'(\eta) = -2E_T\{b\, g'(\ell+\eta)\}$, so $\Delta R'(0) = -2E_T(bw)$ and $|\Delta R''(\eta)| \le 2\sup|g''| < 1/5$. Hence $|\Delta R(\eta) - \Delta R(0) + 2E_T(bw)\eta| \le \gamma^2/10$ for $|\eta| \le \gamma$. Evaluating at $\eta = \pm\gamma$ and bounding the remaining $\eta$ gives $\max_{|\eta|\le\gamma}\Delta R(\eta) = \Delta R(0) + 2\gamma|E_T(bw)| + O(\gamma^2)$ and $\min_{|\eta|\le\gamma}\Delta R(\eta) = \Delta R(0) - 2\gamma|E_T(bw)| + O(\gamma^2)$. Therefore $|W_{\mathrm{scal}}(\gamma) - 4\gamma|E_T(bw)| \quad| \le \gamma^2/5$.

**Step 3.** Dividing,

$$\frac{W_{\mathrm{scal}}(\gamma)}{W_{\mathrm{dir}}(\gamma)} = \frac{4\gamma|E_T(bw)| + O(\gamma^2)}{4\gamma E_T(|b|w) + O(\gamma^3)} \to \kappa. \qquad \square$$

**Remarks.** (i) By $|E_T(bw)| \le E_T(|b|w)$, $\kappa \in [0,1]$. $\kappa = 1$ if and only if $b(Z)$ has $P_T$-almost surely constant sign on $\{w(Z) > 0\}$, in which case $\Delta R(\eta)$ is monotone and the corner tilts $\eta = \pm\gamma$ attain the direct endpoints, so $W_{\mathrm{scal}} = W_{\mathrm{dir}}$ for every $\gamma$. (ii) If $E_T(bw) = 0$, then $W_{\mathrm{scal}}(\gamma) = O(\gamma^2)$ and the ratio tends to zero. (iii) $\kappa$ is a first-order quantity. At $\gamma = 0.2$ the ratio and $\kappa$ differed by at most $0.002$, and at $\gamma = 0.5$ by at most $0.021$, in the analyzed settings (Web Table 1). (iv) The plug-in $\hat{\kappa}$ replaces $m_S$ by its IPCW estimate and $E_T$ by the target empirical mean, and therefore uses no target outcomes. (v) Because $Z$-conditional constant tilts with $|\eta| \leq \gamma$ belong to the profile-varying class, $W_{\text{scal}}(\gamma) \leq W_{\text{dir}}(\gamma)$ for every $\gamma$; $\kappa$ quantifies how much of the direct width this homogeneity restriction removes.

**Web Appendix A.7. One-Step Endpoint Estimator (Proposition 4)**

Fix $\gamma \in \mathcal{G}$ and an endpoint $E \in \{L, U\}$. Define the corner sign

$$s_L(z) = \begin{cases} 1, & b(z) \geq 0, \\ -1, & b(z) < 0, \end{cases} \qquad s_U(z) = -s_L(z),$$

and write $s = s_E$. With $g = \text{expit}$, define

$$f(z, m) = a(z) - 2b(z)\, g\{\text{logit} m + s(z)\gamma\}.$$

By Theorem 1,

$$E_\gamma = E_T\{f(Z, m_S(Z))\}.$$

On $m \in [\epsilon_m, 1 - \epsilon_m]$, the map $m \mapsto f(z, m)$ is twice continuously differentiable, with

$$f'(z, m) \equiv \partial_m f(z, m) = \frac{-2b(z)\, g'\{\text{logit} m + s(z)\gamma\}}{m(1 - m)}.$$

Because $|b| \leq 1$, there are constants $C_1$ and $C_2$ such that

$$|f'(z, m)| \leq C_1, \qquad |\partial_m^2 f(z, m)| \leq C_2$$

uniformly in $z$. Let $r = dP_T^Z / dP_S^Z \leq B_0$ (Section 2.5), and let $\omega$ denote the fixed-horizon IPCW weight of Section 2.5 evaluated at the true $G$. Under $T^* \perp C \mid H$ with $H = Z$ and censoring positivity,

$$E_S(\omega \mid Z) = 1, \qquad E_S\{\omega(Y_\tau - m_S(Z)) \mid Z\} = 0.$$

The one-step estimator of $E_\gamma$ is

$$\hat{E} = P_{T,n} f(Z, \hat{m}) + P_{S,n}\{\hat{r}(Z)\, f'(Z, \hat{m})\, \hat{\omega}\, (Y_\tau - \hat{m})\},$$

where $\hat{m}$ and $\hat{\omega}$ are the primary nuisance estimates and $\hat{r}$ is the normalized density-ratio estimate described in Web Appendix C.2.

**Proposition 4.** Assume conditions (R1) and (R3)–(R6), (R9), and (R10) of Web Appendix A.5; in particular, censoring is conditionally independent given $H = Z$, and the Cox–Breslow estimator $\hat{G}$ is correctly specified and asymptotically linear. Assume further that, with probability tending to one,

$$\hat{m} \in [\epsilon_m, 1 - \epsilon_m], \qquad 0 \le \hat{r} \le B,$$

where $B < 20$, so that the density-ratio cap of 20 is inactive with probability tending to one (under (b), this requires $B_0 < 20$), and that $\hat{m}$ and $\hat{r}$ converge in $L_2(P_S)$ to limits $\tilde{m}$ and $\tilde{r}$ and lie in $P_S$-Donsker classes, as for the parametric models used here; otherwise, the correction term is cross-fitted.

In probability, $\hat{E} \to E_\gamma + B_{os}$, where

$$B_{os} = E_T[f(Z,\tilde{m}) - f(Z,m_S) - f'(Z,\tilde{m})\{\tilde{m} - m_S\}] + E_S[\{\tilde{r} - r\}f'(Z,\tilde{m})\{m_S - \tilde{m}\}].$$

Consequently,

$$|B_{os}| \le \frac{C_2 B_0}{2} \parallel \tilde{m} - m_S \parallel_{2,S}^2 + C_1 \parallel \tilde{r} - r \parallel_{2,S} \parallel \tilde{m} - m_S \parallel_{2,S}.$$

The plug-in estimator instead converges to

$$E_\gamma + E_T\{f(Z,\tilde{m}) - f(Z,m_S)\},$$

whose bias is first order in $\tilde{m} - m_S$.

Suppose $\tilde{m} = m_S$, $\tilde{r} = r$, and

$$\parallel \hat{m} - m_S \parallel_{2,S} = o_p(n^{-1/4}), \qquad \parallel \hat{r} - r \parallel_{2,S} \parallel \hat{m} - m_S \parallel_{2,S} = o_p(n^{-1/2}).$$

Then

$$\hat{E} - E_\gamma = P_{T,n}\phi_T + P_{S,n}\phi_S + \Gamma_{os}(\hat{G} - G) + o_p(n^{-1/2}),$$

where

$$\phi_T = f(Z,m_S) - E_\gamma, \qquad \phi_S = r(Z)\, f'(Z,m_S)\, \omega\, (Y_\tau - m_S),$$

and $\Gamma_{os}$ is the linearization, through the Cox–Breslow influence function of Web Appendix A.5, of the map

$$G^{\dagger} \mapsto P_S\{r\, f'(Z, m_S)\, (\omega_{G^{\dagger}} - \omega_G)(Y_\tau - m_S)\},$$

with $\omega_{G^{\dagger}}$ the IPCW weight computed from $G^{\dagger}$. This censoring term generally differs from that of the plug-in estimator in Web Appendix A.5: the dependence of $\hat{m}$ on $\hat{G}$ has no first-order effect here, because the one-step estimator is first-order insensitive to errors in $\hat{m}$ when $\tilde{r} = r$. Hence the stacked one-step endpoints are jointly asymptotically linear, and the pairs-bootstrap argument of Theorem 2 applies with the parametric nuisance models used here.

**Proof.**

Decomposition. Write

$$\hat{E} - E_\gamma = (P_{T,n} - P_T) f(\hat{m}) + \{P_T f(\hat{m}) - P_T f(m_S)\} + P_{S,n}\{\hat{r} f'(\hat{m}) \hat{\omega} (Y - \hat{m})\}.$$

The last term equals

$$(P_{S,n} - P_S)\{\hat{r} f'(\hat{m}) \omega (Y - \hat{m})\} + P_S\{\hat{r} f'(\hat{m}) \omega (Y - \hat{m})\} + P_{S,n}\{\hat{r} f'(\hat{m})(\hat{\omega} - \omega)(Y - \hat{m})\}.$$

Conditional mean. By the conditional moment identities and $P_S(rh) = P_T h$,

$$P_S\{\hat{r} f'(\hat{m}) \omega (Y - \hat{m})\} = P_T\{f'(\hat{m})(m_S - \hat{m})\} + P_S\{(\hat{r} - r) f'(\hat{m})(m_S - \hat{m})\}.$$

Remainder. Combining (ii) with $P_T f(\hat{m}) - P_T f(m_S)$ leaves

$$R = P_T\{f(\hat{m}) - f(m_S) - f'(\hat{m})(\hat{m} - m_S)\} + P_S\{(\hat{r} - r) f'(\hat{m})(m_S - \hat{m})\}.$$

A second-order Taylor expansion and $dP_T = r\, dP_S$ bound the first term, and the Cauchy–Schwarz inequality bounds the second:

$$|R| \le \frac{C_2 B_0}{2} \parallel \hat{m} - m_S \parallel_{2,S}^2 + C_1 \parallel \hat{r} - r \parallel_{2,S} \parallel \hat{m} - m_S \parallel_{2,S}.$$

Part (a). Replace the estimates by their limits. The empirical-process terms vanish by the Glivenko–Cantelli property, and the term involving $\hat{\omega} - \omega$ vanishes by uniform consistency of $\hat{G}$ and the lower bound on $G$.

Part (b). With $\tilde{r} = r$, the second term of $R$ involves $\hat{r} - r$, so the rate conditions give $R = o_p(n^{-1/2})$. Asymptotic equicontinuity (Donsker classes or cross-fitting) and $L_2$ convergence give

$$(P_{T,n}-P_T)f(\widehat{m}) = (P_{T,n}-P_T)f(m_S) + o_p(n^{-1/2}),$$

$$(P_{S,n}-P_S)\{\hat{r}f'(\widehat{m})\omega(Y-\widehat{m})\} = (P_{S,n}-P_S)\phi_S + o_p(n^{-1/2}).$$

Finally,

$$P_{S,n}\{\hat{r}f'(\widehat{m})(\widehat{\omega}-\omega)(Y-\widehat{m})\} = P_S\{rf'(m_S)(\widehat{\omega}-\omega)(Y-m_S)\} + o_p(n^{-1/2}),$$

which is a Hadamard-differentiable linear functional of the Cox–Breslow estimator. The functional delta method of Web Appendix A.5 therefore yields $\Gamma_{os}(\hat{G}-G)$. ▫

**Remarks.**

The weighted logistic score equations give $P_{S,n}\{\widehat{\omega}\,\phi(Z)(Y-\widehat{m})\} = 0$. The correction is therefore identically zero when $\hat{r}\,f'(\cdot,\widehat{m})$ lies in the span of $\phi$, so the plug-in estimator is already first-order insensitive to misspecification in those directions.

Because $f$ is nonlinear in $m$, $B_{os} \neq 0$ in general when $\widetilde{m} \neq m_S$, even if $\tilde{r} = r$. The estimator is rate double robust, not doubly robust in the model-specification sense.

The censoring term $\Gamma(\hat{G}-G)$ is not orthogonalized. Augmenting $\omega$ with an outcome model for the event time would remove its first-order dependence on $G$.

In the simulation data-generating mechanism of Web Appendix C.2, $\log r$ is linear in $\operatorname{logit} q_1$, so the density-ratio classifier is correctly specified.

If $\tilde{r} \neq r$, the remainder contains the first-order term

$$P_S\{(\tilde{r}-r)\,f'(Z,m_S)\,(m_S-\widehat{m})\},$$

which is of order $n^{-1/2}$ for the parametric $\widehat{m}$ used here. The expansion in (b) then does not apply, and the first-order insensitivity to outcome-regression error is lost. Because $\hat{E}$ remains a smooth functional of finitely many parametric M-estimators and of $\hat{G}$, it is still asymptotically linear, with an additional term from $\widehat{m}$, and the pairs bootstrap remains valid by the argument of Theorem 2. In the applications, correct specification of the density-ratio model is a separate assumption.

**Numerical results.**

In simulation (Web Appendix C.2; Web Table 16), the population limit of the bias under the omitted nonlinear outcome term was 1.08–1.15 ($\times 10^{-3}$) for the plug-in estimator and −0.02 to 0.02 ($\times 10^{-3}$) for the one-step estimator. At the Rotterdam–GBSG scale, the one-step estimator reduced finite-sample bias from 1.2–2.2 to 0.1–1.2 ($\times 10^{-3}$), and outer coverage was 0.957 for both estimators. At the NSCLC scale, finite-sample bias of 1.8–13.9 ($\times 10^{-3}$) affected both estimators, and one-step coverage (0.844–0.847) did not improve on plug-in coverage (0.860–0.863).

In the applications (Web Table 17), the one-step estimator was computed from the same bootstrap draws and critical-value grid as the reported plug-in analysis. Its identified-set endpoints differed from the plug-in endpoints by at most 0.011 in NSCLC and 0.0014 in Rotterdam–GBSG, and its decisions coincided with the plug-in decisions in all 60 cells. At Rotterdam–GBSG seed 62 and $\gamma = 0.2$, its envelope was [−0.0486, −0.0010]. Density-ratio estimates never reached the cap of 20 at the point estimates (maximum 6.58; at most 4 of 999 bootstrap draws per configuration), and the censoring floor was not activated.

## Web Appendix B. Details of the NSCLC and Rotterdam–GBSG analyses

### Web Appendix B.1. NSCLC source and target cohorts

The CNUHH source data comprised 400 patients with lung cancer whose tumor regions of interest were manually delineated by a nuclear medicine physician, and 179 of these patients were also included in previously reported CNUHH cohorts (Oh et al., 2021; Oh et al., 2023). We excluded 57 patients with small-cell lung cancer, leaving 343 patients with NSCLC and complete clinical covariate data. One further patient had a recorded survival time of zero and was excluded because a positive follow-up time is required for the fixed-horizon outcome, yielding 342 patients with 105 observed deaths. Overall survival in CNUHH was measured from the date of clinical diagnosis. Antecedent CNUHH work reported that pretreatment FDG PET/CT was obtained at approximately the time of diagnosis (Oh et al., 2023). The diagnosis-to-imaging interval, however, was not available for the present cohort.

For each of four prespecified seeds, an event-stratified two-to-one split of the 342 patients produced a predictor-development subset of 228 patients (70 deaths) and a nonoverlapping source-calibration subset of 114 patients (35 deaths). Subset sizes and event counts were identical across seeds, whereas subset membership varied. The development subset was used only to construct and lock both predictors; the calibration subset supplied the right-censored outcomes used to estimate the source conditional survival probability and the censoring model.

The Stanford and VA target cohorts were obtained from the public NSCLC-Radiogenomics collection of The Cancer Imaging Archive (Bakr et al., 2018). Of 88 initially available VA cases, seven were excluded because the harmonized metabolic tumor volume was insufficient for feature extraction and one because PET-mask registration failed, yielding 80 VA patients (32 deaths); the Stanford sample comprised 60 patients (22 deaths). These exclusions reflected imaging-feasibility criteria rather than model predictions or comparative performance. Target overall survival was indexed from the CT scan date. These target cohorts had been used in our antecedent model-development work (Choi, Kim and Kim, 2026). The application is therefore a retrospective emulation: target outcomes were withheld from the finalized outcome-free analysis and were used only in the separate descriptive comparison reported in Web Table 14. Web Table 9 summarizes cohort composition and analytic roles.

CNUHH tumors were manually delineated, whereas target tumors were segmented using nnU-Net (Isensee et al., 2019). To obtain a common imaging representation, metabolic tumor volume was defined in all cohorts using a threshold of 40% of the maximum standardized uptake value. This standardizes the feature definition, but residual differences in image acquisition and segmentation may remain.

**Web Appendix B.2. Locked predictors and prediction horizon**

The reference predictor $q_0$ was a clinical-only Cox proportional hazards model using age, sex, ever-smoking status, log-transformed pack-years, and clinical T, N, and M stage, with a Breslow baseline-hazard estimator fitted in the development subset. The candidate predictor $q_1$ added PET radiomic features from the harmonized metabolic-tumor-volume representation to the same clinical base. The initial radiomic block had 1,112 features. Feature screening and sparse principal component analysis were carried out in a prespecified source-only inductive pipeline (Zou, Hastie and Tibshirani, 2006). A clinical Cox model was fitted first, followed by a ridge-penalized Cox model on the sparse principal components with the clinical linear predictor as a fixed offset; fixed-horizon survival probabilities were obtained from a Breslow baseline estimator fitted in the development subset. All feature selection, dimensionality reduction, tuning, and fitting used source development data only. The four seeds generated four distinct locked configurations through differences in source splitting, feature selection, sparse principal component dimension, and penalization; they are not independent replications, ensemble members, or voting components. Web Table 10 gives the common construction and the seed-specific hyperparameters.

The two-year horizon ($\tau$ = 730 days) was fixed for the finalized outcome-free analysis and was not selected on the basis of the resulting certification decisions; it was guided primarily by the source-calibration information available at the horizon. Across seed-specific calibration subsets, the median observed event-or-censoring time was approximately one year and the maximum approximately 3.8 years. At longer horizons fewer source patients remain under observation, increasing uncertainty in the source outcome model, the Breslow baseline survival function, and the censoring distribution. Prespecified one- and three-year analyses served as horizon-sensitivity analyses; all evaluated cells yielded DEFER, and none was flagged as unstable with respect to censoring positivity (Web Table 12).

**Web Appendix B.3. Nuisance estimation and support diagnostics**

In both applications, the censoring distribution was estimated by an unpenalized conditional Cox model with $H = Z = (q_0, q_1)$ and a Breslow baseline, and the fixed-horizon inverse-probability-of-censoring weights were formed as in Section 2.5 of the main text (Bang and Tsiatis, 2000; Gerds and Schumacher, 2006; Graf et al., 1999). The source conditional survival probability was estimated by unpenalized IPCW-weighted logistic regression on an intercept and the two locked prediction scores standardized in the calibration sample. The standardization parameters, censoring model, and outcome-regression coefficients were estimated from the calibration sample, whereas the locked predictors remained fixed. Estimated censoring probabilities were truncated below at 0.02 as a numerical safeguard, which is distinct from the population-level censoring-positivity assumption. Truncation was not activated in the NSCLC or Rotterdam–GBSG point estimates or bootstrap refits, and no nuisance-fitting failures occurred. An exploratory comparison of alternative censoring estimators under covariate-dependent censoring is reported in Web Table 2; it does not establish robustness to censoring-model misspecification.

Source–target overlap in the NSCLC deployment-summary space was examined with four diagnostics: domain-classification area under the curve (AUC), the proportion of target observations outside the rectangular source calibration range, the importance-weight effective sample size, and the maximum estimated density ratio (Web Table 13). Across the four configurations and both target cohorts, the domain-classification AUC was 0.728–0.783 (Stanford 0.746–0.783; VA 0.728–0.737), 0–1.67% of target observations lay outside the source range, the effective sample size was 87.9%–92.6% of the target sample size, and the maximum density ratio was 2.26–2.65. These diagnostics indicate measurable distributional differences without gross empirical support failure; they cannot verify the transport-positivity assumption.

**Web Appendix B.4. Complete NSCLC results**

All 40 prespecified evaluation cells (two target cohorts, four configurations, five sensitivity values) yielded DEFER because each simultaneous confidence envelope contained zero (Web Table 11). In the seed-42 configuration at $\gamma = 0.2$ displayed in Figure 3 of the main text, the direct identified sets were [+0.0021, +0.0147] in Stanford and [+0.0248, +0.0410] in VA, whereas the separate-risk subtraction intervals were [−0.0203, +0.0371] and [+0.0002, +0.0655]. The direct simultaneous envelopes were [−0.0237, +0.0455] and [−0.0267, +0.0958], and the separate envelopes were [−0.0422, +0.0850] and [−0.0486, +0.1324]. The separate-to-direct width ratios were 4.53 and 4.02 for the identified sets and 1.84 and 1.48 for the envelopes. Because the direct identified sets excluded zero while the envelopes did not, these two configurations were limited by sampling rather than identification uncertainty. Across the full grid, DEFER reflected identification uncertainty, sampling uncertainty, or both, depending on the configuration and sensitivity value.

Target outcomes were examined only after the outcome-free decisions had been fixed (Web Table 14). Averaged over the four configurations, the candidate-minus-reference Brier-risk contrast was approximately −0.0076 in Stanford and +0.0237 in VA. Discrimination was summarized by the C-index (Harrell Jr, Lee and Mark, 1996). The corresponding C-index contrasts were approximately +0.027 and −0.012, respectively. Seed-specific bootstrap intervals generally included zero. These descriptive comparisons were not used to select the sensitivity parameter, construct the identified sets, or determine decisions.

**Web Appendix B.5. Rotterdam–GBSG analysis**

We used node-positive patients from the Rotterdam and GBSG breast-cancer cohorts distributed with the survival package in R (Royston and Altman, 2013; Sauerbrei et al., 1999). Rotterdam served as the source ($n$ = 1,546), with nonoverlapping predictor-development ($n$ = 1,030) and calibration ($n$ = 516) subsets, and GBSG as the target ($n$ = 686). The endpoint was five-year recurrence-free survival ($\tau$ = 1,826 days), with follow-up administratively truncated at seven years. The reference Cox predictor included age, menopausal status, tumor size, and hormonal treatment; the candidate additionally included estrogen-receptor status and the transformed positive lymph-node count nodes$^{-1/2}$. These additions were motivated by established prognostic knowledge and were specified without examining GBSG outcomes. Four prespecified seeds defined four distinct locked predictor configurations. Nuisance estimation, the sensitivity grid {0, 0.1, 0.2, 0.3, 0.5}, and the joint pairs bootstrap with $B$ = 999 replicates were as in the NSCLC analysis.

Web Table 15 reports the identified sets, simultaneous envelopes, and decisions for all 20 configuration-by-sensitivity cells. Direct identification yielded CANDIDATE SUPPORTED in all four configurations at $\gamma = 0$, in two at $\gamma = 0.1$, in one at $\gamma = 0.2$, and in none at $\gamma = 0.3$ or 0.5. Separate-risk subtraction gave the same decision in 19 cells; at seed 62 and $\gamma = 0.2$ the direct envelope was [−0.0486, −0.0017] (CANDIDATE SUPPORTED), whereas the separate envelope was [−0.0586, +0.0096] (DEFER). In a separate descriptive GBSG outcome analysis, the candidate-minus-reference Brier-risk contrast was −0.0097 (bootstrap interval containing zero) and the C-index contrast was +0.081 (interval excluding zero); these outcomes did not determine the outcome-free decisions.

## Web Appendix C. Simulation details

### Web Appendix C.1. Aims

The simulation study was organized according to the aims, data-generating mechanisms, estimands, methods, and performance measures framework (Morris, White and Crowther, 2019). It had four aims. Aim 1 characterized the direct-versus-separate identification geometry of Proposition 2, including how the positions of $q_0$ and $q_1$ relative to one half affect identified-set width. Aim 2 examined finite-sample performance of the bootstrap envelopes, distinguishing $\gamma$-grid outer-envelope coverage, $\gamma$-grid two-sided endpoint coverage, and per-$\gamma$ true-contrast containment. Aim 3 evaluated CANDIDATE SUPPORTED, REFERENCE SUPPORTED, and DEFER frequencies, wrong-direction certification, and robustness-value recovery under known decision regimes. Aim 4 evaluated stress conditions, including limited source–target overlap, source outcome-model misspecification, shifts beyond the prespecified sensitivity class, and finite-sample behavior of the primary conditional Cox–Breslow procedure across source information levels.

### Web Appendix C.2. Data-generating mechanisms

Controlled-geometry illustration. At $\gamma = 0.3$, using one numerical sample of 300,000 target profiles, we varied the proportion of profiles with $q_0$ and $q_1$ on the same side of one half while holding the remaining construction fixed. This illustration isolates the strictness condition of Proposition 2 and is not a Monte Carlo coverage study (Web Table 3; Web Figure 1A).

Application-scale comparator and mirror study. Source calibration used $n_S = 114$ and target sizes $n_T \in \{60, 80\}$, matching the NSCLC application. Seven principal scenarios covered no conditional outcome shift, a constant shift, an in-class profile-varying shift, a sign-changing shift, limited source–target overlap, source outcome-model misspecification, and an out-of-class shift; candidate-superior mirror versions of the no-shift and profile-varying scenarios assessed the opposite decision direction. The sensitivity grid was $\{0, 0.1, 0.2, 0.3, 0.5\}$, with $\gamma = 0$ serving as the conditional-transportability benchmark. This component used Gaussian source and target covariates, fixed prediction functions, a logistic source conditional survival law, scenario-specific log-odds shifts, Weibull event times, and covariate-independent censoring, with inverse-probability-of-censoring weights based on the marginal Kaplan–Meier estimator (Kaplan and Meier, 1958). Limited overlap arose from shifting the target covariate distribution; misspecification added an omitted nonlinear term while keeping the working linear outcome model; in the out-of-class scenario the true conditional shift exceeded the reported sensitivity class. Each condition used 1,000 Monte Carlo replications and $B = 199$ bootstrap replicates to limit computation across the full design (Web Table 3). Targeted primary-implementation study. Because the comparator used covariate-independent censoring and marginal Kaplan–Meier estimation, we separately evaluated the conditional Cox–Breslow procedure used in the applications. The NSCLC-scale arms used $n_S = 114, n_T \in \{60, 80\}$, and approximately 65% censoring; a second arm used $n_S = 516, n_T = 686$, and approximately 38% censoring to match the Rotterdam–GBSG information scale. Both used conditional Cox–Breslow censoring estimation, IPCW-weighted logistic estimation of the source conditional survival probability, and the joint pairs bootstrap of Section 2.6 of the main text. A focused factorial analysis used the same data-generating mechanism with $n_T = 80$ fixed, varying only the source calibration size ($n_S = 114 \; or \; 516$) and the censoring level (approximately 65%, 50%, or 38%); each cell used 1,000 Monte Carlo replications and $B = 199$ bootstrap replicates (Web Table 8). After these six cells had been run, source sizes $n_S \in \{200, 300, 400\}$ were added under a protocol fixed before the additional cells were run, with the same code, data-generating mechanism, and seeding rule; all 15 cells are reported (Web Table 8; Web Figure 3). These studies characterize finite-sample behavior at application-relevant information levels and do not modify the asymptotic result of Theorem 2.

Decision-regime study. Four data-generating mechanisms defined five operational regimes: candidate-superior, reference-superior, ambiguous, finite crossing, and no crossing within range. The last regime used the same data-generating mechanism as the finite-crossing regime but truncated the robustness-value search at $\gamma = 0.5$, placing the population crossing outside the examined range. The five regimes were crossed with two censoring mechanisms, yielding 10 conditions, with $n_S = 3{,}000, n_T = 1{,}500$, 1,000 Monte Carlo replications, and $B = 999$ bootstrap replicates to reduce nuisance-estimation instability and isolate decision behavior. Simultaneous inference used a grid with spacing 0.02 up to a regime-specific upper bound of 2.0 (0.5 for the no-crossing-within-range regime), and decisions were evaluated at $\gamma = 0.3$. Event times followed a Weibull model with shape 1.5; censoring was either covariate-independent exponential with marginal Kaplan–Meier estimation or covariate-dependent exponential with Kaplan–Meier estimation within prespecified covariate strata. This study did not use the primary conditional Cox–Breslow censoring estimator (Web Table 3).

Across components, $q_0$ and $q_1$ were prespecified fixed prediction functions that were never re-estimated or recalibrated with target outcomes. Population truth, including the true contrast and the sharp endpoints, was computed from the known target law, using 120-node Gauss–Hermite quadrature where numerical integration was required. Simulated target outcomes defined the truth only and were never supplied to the estimator. Scenarios, sensitivity grids, decision thresholds, and analysis procedures were fixed before the final Monte Carlo summaries were compiled.

One-step comparison. The primary-implementation data-generating mechanism was used at two information scales:

NSCLC scale: $n_S = 114, n_T = 80$, approximately 65% censoring;

Rotterdam–GBSG scale: $n_S = 516, n_T = 686$, approximately 38% censoring.

Two source outcome laws were considered. Law A was the correctly specified law above. Law B added the term

$$2.5(q_1 - 0.55)^2$$

to the logit of the source outcome law, the omitted-term form used in the outcome-model misspecification scenario of the comparator study. The working outcome model remained logistic in standardized $(q_0, q_1)$. The density ratio was estimated by an unpenalized logistic domain classifier on standardized $(logitq_0, logitq_1)$, pooled over source and target, normalized to mean one over the source sample, and capped at 20. Plug-in and one-step endpoints were computed from the same data and the same $B = 199$ bootstrap draws, with method-specific sup-t critical values over $\mathcal{G}$. Each of the four conditions used 1,000 Monte Carlo replications, and population truth was computed from 2,000,000 target draws.

**Web Appendix C.3. Estimands, comparisons, and performance measures**

Each simulation targeted the true Brier-risk contrast, the sharp sensitivity-indexed endpoints, and, where relevant, the direction-specific robustness value. Direct identification and separate-risk subtraction were applied to the same simulated samples with the same nuisance-estimation and bootstrap procedures, so that differences reflected only whether the shared unidentified target outcome law was preserved. Two secondary benchmarks were considered: the conditional-transportability benchmark $\gamma = 0$ and a scalar constant-shift restriction on the same deployment summary (Web Table 1). Because the scalar model imposes stronger homogeneity than the primary profile-varying class, it illustrates the effect of restricting shift heterogeneity rather than an efficiency comparison under the same sensitivity model. For binary outcomes, the scalar constant-shift model is the Z-conditional analogue of the exponential-tilt sensitivity model of Steingrimsson et al. (2024); it permits a direct sensitivity analysis of the Brier-risk contrast but restricts the conditional log-odds departure to be constant across deployment profiles.

Identification performance was summarized by identified-set widths and the separate-to-direct width ratio. The targeted primary-implementation study reported $\gamma$-grid outer-envelope coverage, the event targeted by Theorem 2, together with the more demanding $\gamma$-grid two-sided endpoint coverage. The decision-regime study reported $\gamma$-grid two-sided endpoint coverage. The application-scale comparator reported per-$\gamma$ containment of the single true contrast by the direct simultaneous band; this event does not assess endpoint calibration. In the source outcome-model misspecification scenario, we also recorded whether the plug-in identified set contained the true contrast. For the decision-regime study, we recorded decision frequencies, wrong-direction certifications, robustness-value recovery, nuisance-fitting or bootstrap-refitting failures, and activation of the 0.02 censoring-probability floor. Proportions were accompanied by Monte Carlo standard errors or Wilson binomial confidence limits (Wilson, 1927). When no wrong-direction certification occurred, an exact one-sided binomial upper confidence bound was reported (Clopper and Pearson, 1934).

**Web Appendix C.4. Additional results**

Identification geometry. In the controlled-geometry illustration, the separate-to-direct identified-set width ratio increased from 1.00 to 5.73 as the proportion of target profiles with both predictions on the same side of one half rose from 0 to 1, and equaled 1.00 when the predictions straddled one half (Web Figure 1A). In the application-scale comparator, direct identification was narrower in every scenario-by-target-size cell, with identified-set width ratios of 3.43–3.63 and envelope width ratios of 1.47–1.59 at $\gamma = 0.2$ (Web Table 4; Web Figure 1B).

*Comparator containment.* Per-$\gamma$ true-contrast containment by the direct simultaneous band was approximately 0.97–0.98 under standard conditions and 0.891 at $\gamma = 0.2$ with limited overlap for $n_T = 80$ (Web Tables 4 and 5; Web Figure 1C). Because a wide identified set can make point containment easy, these rates should not be read as endpoint calibration; similarly, the misspecification value of 0.968 does not establish robustness to source outcome-model misspecification. Under deliberate misspecification, plug-in identified-set containment varied substantially across working models (Web Table 6). Under the out-of-class shift, per-$\gamma$ containment by the direct band was approximately 0.91–0.98 even though the true target law lay outside the reported class (Web Table 5), so such containment validates neither the sensitivity model nor endpoint calibration.

**Primary implementation.** At the NSCLC scale, outer-envelope coverage was 0.861 and 0.847 for $n_T = 60$ $and$ 80, and at the Rotterdam–GBSG scale it was 0.946 (Table 1 of the main text). The outer-envelope and two-sided events differed in only 2, 0, and 3 of 1,000 replications in the three settings, so the undercoverage at the NSCLC scale is not attributable to the stronger two-sided criterion. No fatal fitting failures occurred; numerical nonconvergence or activation of the 0.02 floor was rare at the NSCLC scale and absent or negligible at the Rotterdam–GBSG scale. In the factorial analysis (Web Table 8; Web Figure 3), coverage generally increased with source size and decreased with censoring, but not monotonically: at 65% censoring it was 0.914, 0.907, and 0.920 for $n_S = 300, 400$ and 516, and at 38% censoring 0.947 for $n_S = 400$ and 0.940 for $n_S = 516$. Wilson intervals included 0.95 in the six cells with $n_S \geq 300$ and at most 50% censoring (point estimates 0.938–0.950) and in no other cell, although the upper limit for $n_S = 300$ with 50% censoring was only 0.951. At similar numbers of source events (about 180–186), coverage was 0.920 with $n_S = 516$ and 65% censoring but 0.947 with $n_S = 300$ and 38% censoring. Mean sup-t critical values decreased from 2.25 to 2.09 as source information increased. The factorial value of 0.859 and the value 0.847 in Table 1 of the main text, for comparable settings, arose from independent Monte Carlo runs and differ within Monte Carlo error.

**Finite-sample coverage diagnosis.** To investigate the observed coverage shortfall, we further examined two factorial cells using their existing Monte Carlo replications and seeds: $n_S = 114$ with 65% censoring (coverage 0.859) and $n_S = 516$ with 50% censoring (coverage 0.950). The recomputed envelopes and critical values reproduced the stored results exactly. A replication was classified as missing at the lower (upper) side if its lower (upper) confidence limit failed to cover the true lower (upper) endpoint at one or more grid values of $\gamma$. In the first cell, 121 of the 141 non-covering replications missed only at the lower side, 20 only at the upper side, and none at both. Plug-in lower endpoints were biased upward by 0.008–0.013 across $\gamma$ (0.26–0.41 Monte Carlo standard deviations); recomputing the target-sample endpoints with the true source survival probability in place of its estimate reduced this bias to less than 0.0002, suggesting that estimation of $m_S$ from about 40 source events contributes to it. Bootstrap standard errors averaged 0.86–0.88 of the Monte Carlo standard deviations, and their correlation with the signed estimation error was −0.17 for the lower endpoint at $\gamma = 0.2$. Two oracle calculations, which use Monte Carlo quantities unavailable in practice, gave coverage 0.944 when each replication's bootstrap standard errors were replaced by the Monte Carlo standard deviations while its own bootstrap critical value was retained, and 0.887 when the Monte Carlo mean bias at each $\gamma$ and endpoint was subtracted from both limits with the bootstrap standard errors and critical values unchanged. In the second cell, the standard-error ratio was 0.99, the bias was at most 0.18 Monte Carlo standard deviations, misses were split evenly (25 lower only, 25 upper only, none at both), and the correlation was 0.48. These exploratory diagnostics suggest that upward endpoint bias and underestimated bootstrap standard errors contribute to the shortfall in small, heavily censored calibration samples; they do not constitute a finite-sample correction.

**Decisions and robustness values.** In the decision-regime study, $\gamma$-grid two-sided endpoint coverage ranged from 0.939 to 0.950 across the 10 conditions. At $\gamma = 0.3$, the candidate-superior regime yielded CANDIDATE SUPPORTED in 41.9% and 35.5% of replications under independent and covariate-dependent censoring, with DEFER otherwise; the reference-superior regime yielded REFERENCE SUPPORTED in 77.1% and 73.0%; the ambiguous regime yielded DEFER throughout; and the finite-crossing and no-crossing-within-range regimes yielded CANDIDATE SUPPORTED in all replications (Web Table 7; Web Figure 2). Root mean squared errors of the robustness value were 0.051, 0.031, and 0.055 for the candidate-superior, ambiguous, and finite-crossing regimes; in the reference-superior regime, the population reference-direction robustness value was 0.44, with root mean squared errors of 0.049 and 0.052 under the two censoring mechanisms. In the candidate-superior mirror scenarios of the comparator, no wrong-direction REFERENCE SUPPORTED decision occurred in 1,000 replications per condition, giving an exact one-sided 95% upper bound of approximately 0.003 (Web Table 7).

**Web Table 1.** Scalar constant-shift versus profile-varying sensitivity models: widths at $\gamma = 0.2$, first-order diagnostic $\kappa$, and population robustness values

| Setting | Seed | $P_T(q_1 > q_0)$ | Profile-varying width | Scalar width | Width ratio | $\kappa$ | Favored at $\gamma = 0$ | Robustness value: profile-varying/scalar |
|---|---|---|---|---|---|---|---|---|
| NSCLC–Stanford | 42 | 0.95 | 0.01269 | 0.01203 | 0.948 | 0.948 | Reference | 0.270/0.285 |
| | 52 | 0.92 | 0.00859 | 0.00659 | 0.767 | 0.768 | Reference | 0.755/1.110 |
| | 62 | 0.60 | 0.00901 | 0.00200 | 0.222 | 0.223 | Candidate | 0.340/1.405 |
| | 72 | 0.90 | 0.00684 | 0.00463 | 0.677 | 0.677 | Candidate | 0.410/0.550 |
| NSCLC–VA | 42 | 0.90 | 0.01622 | 0.01266 | 0.780 | 0.780 | Reference | 0.980/1.700 |
| | 52 | 0.90 | 0.01246 | 0.00744 | 0.597 | 0.597 | Reference | 0.885/2.365 |
| | 62 | 0.79 | 0.00891 | 0.00070 | 0.079 | 0.081 | Candidate | 0.470/1.775 |
| | 72 | 0.88 | 0.00770 | 0.00436 | 0.565 | 0.565 | Candidate | 0.570/0.825 |
| Rotterdam–GBSG | 42 | 0.47 | 0.02094 | 0.00158 | 0.075 | 0.075 | Candidate | 0.455/1.750 |
| | 52 | 0.47 | 0.02279 | 0.00215 | 0.094 | 0.094 | Candidate | 0.335/1.420 |
| | 62 | 0.46 | 0.02091 | 0.00105 | 0.050 | 0.049 | Candidate | 0.470/1.915 |
| | 72 | 0.46 | 0.02320 | 0.00155 | 0.067 | 0.067 | Candidate | 0.285/1.385 |

Notes: Both sensitivity formulations condition on the same deployment summary $Z = (q_0, q_1)$ and use the same IPCW-weighted logistic estimate of $m_S$. NSCLC rows use the primary SparsePCA candidate. Rotterdam–GBSG rows use the literature-motivated candidate. The scalar constant-shift model imposes $\delta(z) \equiv \eta$ with $|\eta| \leq \gamma$, whereas the profile-varying model permits any $\delta(z)$ with $|\delta(z)| \leq \gamma$. Because the scalar class is contained in the profile-varying class, it cannot yield a wider set. A narrower scalar range reflects its homogeneity restriction rather than an efficiency gain. $\kappa = |E_T(bw)|/E_T(|b|w)$, with $w = m_S(1 - m_S)$, is the first-order $(\gamma \to 0)$ limit of the width ratio (Web Appendix

A.6). The scalar range was computed over a 401-point $\eta$ grid on $[-\gamma, \gamma]$. For Rotterdam–GBSG seed 62, $\Delta R(\eta)$ was not monotone in $\eta$, and the range was taken over the grid. Robustness values are population plug-in quantities for the direction favored at $\gamma = 0$ (Proposition 3), computed on a grid with step 0.005, and do not reflect sampling uncertainty. All quantities use source calibration data and unlabeled target predictions only. NSCLC rows use the CT-date-indexed cohorts (Stanford $n = 60$, VA $n = 80$).

**Web Table 2.** Exploratory sensitivity to alternative censoring estimators under covariate-dependent censoring

| $\psi$ | $\gamma$ | Marginal-KM identified set | Conditional-KM identified set | $\lvert\Delta U\rvert$ | Decision flip |
|---|---|---|---|---|---|
| 0.0 | 0.2 | [-0.0095, +0.0003] | [-0.0095, +0.0004] | 0.0001 | No |
| 0.5 | 0.2 | [-0.0136, -0.0046] | [-0.0137, -0.0048] | 0.0002 | No |
| 1.0 | 0.2 | [-0.0169, -0.0085] | [-0.0177, -0.0097] | 0.0012 | No |
| 1.0 | 0.5 | [-0.0223, -0.0016] | [-0.0230, -0.0030] | 0.0014 | No |

Notes: This exploratory sensitivity analysis used $n_S = 114, n_T = 40, \tau = 1$, and 400 Monte Carlo replications, with ~54% censoring. Censoring times followed $C \sim Exp\{\lambda exp(\psi X)\}$, with increasing $\psi$ representing stronger covariate-dependent censoring. The comparison contrasts marginal Kaplan–Meier IPCW with a simple conditional Kaplan–Meier IPCW procedure. No decision flip was observed in the reported settings, and the largest upper-endpoint difference was 0.0014. This analysis is an empirical stress test only and does not establish robustness to censoring-model misspecification. Finite-sample performance of the primary conditional Cox–Breslow implementation is evaluated separately in Table 1 of the main text and Web Table 8.

**Web Table 3.** Data-generating mechanisms and key simulation settings

(a) Controlled-geometry illustration

| **Component** | **Specification** |
|---|---|
| Geometry index | $f = P_T\{(q_1 - 1/2)(q_0 - 1/2) > 0\}$ |
| Geometry grid | $f \in \{0, 0.25, 0.50, 0.75, 1.00\}$ |
| Design | Only the relative positions of $q_0$ and $q_1$ with respect to 1/2 were varied; all remaining data-generating components were held fixed. |

Notes: The sensitivity value was $\gamma = 0.3$. This panel used a single numerical sample of 300,000 target profiles to illustrate the identification geometry in Proposition 2 and was not a Monte Carlo coverage study. Only the relative positions of $q_0$ and $q_1$ with respect to 1/2 were varied; all remaining data-generating components were held fixed.

(b) Application-scale comparator and mirror study

| **Component** | **Specification** |
|---|---|
| Sample sizes | $n_S = 114, n_T = \{60,80\}$ |
| Covariate distributions | $X_S \sim N(0,1)$; $X_T \sim N(\mu_T, 1)$, with $\mu_T = 1.20$ for the limited-overlap scenario and 0 otherwise |
| Locked predictors | $q_0(x) = expit(0.10 + 0.50x), q_1(x) = expit(0.35 + 0.90x)$ |
| Source outcome law | $m_S(x) = expit\{-0.25 + 1.40q_0(x) - 0.80q_1(x)\}$ |
| Source-model misspecification | The true source logit additionally contained $2.5\{q_1(x) - 0.55\}^2$, while the fitted working model remained linear in $(q_0, q_1)$ |
| Target outcome shifts | No shift: $\delta(x) = 0$; constant: $\delta(x) = 0.20$; profile-varying: $\delta(x) = 0.20tanh(x)$; sign-changing: $\delta(x) = 0.40sin(2x)$; limited overlap: $\mu_T = 1.20, \delta(x) = 0.20tanh(x)$; $m_S$ misspecification: $\delta(x) = 0.20tanh(x)$, together with the nonlinear source term specified above; out-of-class: $\delta(x) = 0.80tanh(x)$ |
| Event/censoring model | $\tau = 1$; Weibull event-time model with shape 1.5; independent exponential censoring estimated by marginal Kaplan–Meier |
| Sensitivity grid | $\gamma \in \{0.1,0.2,0.3,0.5\}$; $\gamma = 0$ used as the conditional-transportability benchmark |
| Monte Carlo design | 1000 Monte Carlo replications; $B = 199$ joint pairs-bootstrap replications |
| Mirror scenarios | Candidate-superior mirror versions of the no-shift and profile-varying-shift scenarios |

Notes: Negative values of the target Brier-risk contrast favor the candidate predictor. The value $\gamma = 0$ was used only as the conditional-transportability benchmark and was not

included in the eligible-$\gamma$ summaries in Web Table 5. Candidate-superior mirror scenarios were evaluated for the no-shift and profile-varying-shift scenarios.

(c) Decision-regime study

| **Regime** | $i_0$ | $s_0$ | $i_1$ | $s_1$ | $g_{max}$ |
|---|---|---|---|---|---|
| Candidate superior | 0.8 | 0.7 | 0.4 | 0.7 | 2.0 |
| Reference superior | 0.4 | 0.7 | 0.9 | 0.7 | 2.0 |
| Ambiguous | -0.1 | 0.7 | 0.5 | 0.7 | 2.0 |
| Finite crossing | 2.0 | 0.7 | 0.2 | 0.7 | 2.0 |
| No crossing within range | 2.0 | 0.7 | 0.2 | 0.7 | 0.5 |

Notes: Common settings were $n_S = 3000, n_T = 1500, \tau = 1, m_S(x) = expit(0.2 + 0.7x)$, and $q_j(x) = expit(i_j + s_j x)$. The true target shift was $\delta(x) = clip(0.15 + 0.6x, -0.25, 0.25)$, so the true target law entered the reported sensitivity class for $\gamma \geq 0.25$, where $clip(u, a, b) = min\{max(u, a), b)$. The sensitivity grid had spacing 0.02, decisions were evaluated at $\gamma = 0.3$, and each condition used 1,000 Monte Carlo replications with $B = 999$ joint pairs-bootstrap replications. Negative values of the target Brier-risk contrast favor the candidate predictor. The finite-crossing and no-crossing-within-range regimes used the same data-generating mechanism and locked predictors and differed only in the prespecified sensitivity search range, with $\gamma_{max} = 2.0$ and $\gamma_{max} = 0.5$, respectively. Because the no-crossing-within-range regime used $\gamma_{max} = 0.5$, a robustness value exceeding this range is reported as >0.50 rather than as a finite crossing.

**Web Table 4.** Application-scale marginal-Kaplan–Meier comparator: identified-set and confidence-envelope widths and true-contrast containment at $\gamma = 0.2$

| Scenario | $n_T$ | Direct identified-set width | Separate identified-set width | Width ratio | Direct envelope width | Separate envelope width | Width ratio | Direct-band true-contrast containment |
|---|---|---|---|---|---|---|---|---|
| No shift | | 0.0134 | 0.0462 | 3.44 | 0.0845 | 0.1342 | 1.59 | 0.980 |
| Profile-varying shift (in class) | | 0.0134 | 0.0461 | 3.44 | 0.0849 | 0.1346 | 1.59 | 0.982 |
| Limited overlap | 80 | 0.0180 | 0.0651 | 3.62 | 0.1873 | 0.2760 | 1.47 | 0.891 |
| $m_S$ misspecification | | 0.0134 | 0.0459 | 3.43 | 0.0851 | 0.1346 | 1.58 | 0.968 |
| Out-of-class shift | | 0.0134 | 0.0462 | 3.44 | 0.0842 | 0.1340 | 1.59 | - |
| No shift | | 0.0134 | 0.0460 | 3.43 | 0.0859 | 0.1353 | 1.58 | 0.981 |
| Profile-varying shift (in class) | | 0.0134 | 0.0462 | 3.44 | 0.0868 | 0.1368 | 1.58 | 0.981 |
| Limited overlap | 60 | 0.0181 | 0.0658 | 3.63 | 0.1907 | 0.2809 | 1.47 | 0.911 |
| $m_S$ misspecification | | 0.0133 | 0.0457 | 3.43 | 0.0866 | 0.1361 | 1.57 | 0.979 |
| Out-of-class shift | | 0.0134 | 0.0461 | 3.45 | 0.0861 | 0.1358 | 1.58 | - |

Notes: Identified-set widths, confidence-envelope widths, and true-contrast containment rates are reported at $\gamma = 0.2$. The width ratio is defined as the separate-risk identified-set width divided by the direct identified-set width. Direct-band true-contrast containment is the proportion of Monte Carlo replications in which the direct simultaneous band $[\underline{L}_\gamma, \overline{U}_\gamma]$ contains the true target Brier-risk contrast $\Delta R_T$ at the evaluated $\gamma$. This is a per-$\gamma$ point-containment measure, not $\gamma$ -grid endpoint coverage, and should not be interpreted as calibration of the sensitivity-defined endpoints. The $m_S$-misspecification containment values likewise do not establish robustness to outcome-model misspecification. Each scenario used 1000 Monte Carlo replications and 199 bootstrap replications per Monte Carlo sample, with $n_S = 114$. For the out-of-class scenario, a dash is retained in this summary; descriptive per-$\gamma$ point-containment results are reported in Web Table 5(b) and do not validate the sensitivity class.

**Web Table 5.** Application-scale simulation results across sensitivity and stress conditions

(a) Per-$\gamma$ true-contrast containment, confidence-envelope width, and DEFER frequency across eligible sensitivity values

| $n_T$ | Scenario | Eligible $\gamma$ values | Minimum true-contrast containment | Mean true-contrast containment | Mean envelope width | Mean DEFER frequency |
|---|---|---|---|---|---|---|
| 60 | No shift | 0.1, 0.2, 0.3, 0.5 | 0.967 | 0.982 | 0.091 | 0.871 |
| | Constant shift | 0.2, 0.3, 0.5 | 0.974 | 0.983 | 0.094 | 0.899 |
| | Profile-varying shift | 0.2, 0.3, 0.5 | 0.981 | 0.987 | 0.095 | 0.910 |
| | Sign-changing shift | 0.5 | 0.991 | 0.991 | 0.105 | 0.955 |
| | Limited overlap | 0.2, 0.3, 0.5 | 0.911 | 0.921 | 0.202 | 0.832 |
| | $m_S$ misspecification | - | - | - | - | - |
| | Out-of-class shift | - | - | - | - | - |
| 80 | No shift | 0.1, 0.2, 0.3, 0.5 | 0.966 | 0.980 | 0.089 | 0.863 |
| | Constant shift | 0.2, 0.3, 0.5 | 0.982 | 0.988 | 0.092 | 0.909 |
| | Profile-varying shift | 0.2, 0.3, 0.5 | 0.982 | 0.988 | 0.093 | 0.887 |
| | Sign-changing shift | 0.5 | 0.993 | 0.993 | 0.104 | 0.969 |
| | Limited overlap | 0.2, 0.3, 0.5 | 0.891 | 0.906 | 0.199 | 0.809 |
| | $m_S$ misspecification | - | - | - | - | - |
| | Out-of-class shift | - | - | - | - | - |

Notes: Summary quantities are calculated over the listed eligible $\gamma$ values. True-contrast containment is the per-$\gamma$ proportion of Monte Carlo replications in which the direct simultaneous band contains the true target Brier-risk contrast; it is distinct from $\gamma$-grid two-sided endpoint coverage and from outer-envelope containment of the sensitivity-defined identified set. Mean envelope width and mean DEFER frequency summarize the corresponding direct simultaneous-envelope width and DEFER frequency across the eligible $\gamma$ values. The source-outcome-model misspecification scenario is summarized separately in Web Table 6; the corresponding per-$\gamma$ direct-band true-contrast containment values reported in Web Table 4 do not establish robustness to source outcome-model misspecification.

(b) Per-$\gamma$ true-contrast containment under the out-of-class shift

| $\boldsymbol{n_T}$ | $\boldsymbol{\gamma = 0.1}$ | $\boldsymbol{\gamma = 0.2}$ | $\boldsymbol{\gamma = 0.3}$ | $\boldsymbol{\gamma = 0.5}$ |
|---|---|---|---|---|
| 60 | 0.911 | 0.943 | 0.959 | 0.982 |
| 80 | 0.914 | 0.941 | 0.960 | 0.985 |

Notes: The true target law lies outside the prespecified sensitivity class at every reported $\gamma$. The resulting point-containment rates are therefore descriptive and should not be interpreted as validation of the sensitivity assumption or as endpoint-calibration coverage.

**Web Table 6.** Mean plug-in identified-set true-contrast containment

| Estimator | Mean plug-in identified-set true-contrast containment | Mean identified-set width | Mean DEFER frequency |
|---|---|---|---|
| Spline | 0.155 | 0.0109 | 0.091 |
| Enriched | 0.228 | 0.0143 | 0.124 |
| Linear | 0.263 | 0.0165 | 0.147 |
| Ridge | 0.529 | 0.0194 | 0.114 |

Notes: True-contrast containment in this table is the proportion of Monte Carlo replications for which the plug-in identified set contains the true target Brier-risk contrast. This differs from the per- $\gamma$ true-contrast containment by the direct simultaneous band reported in Web Tables 4 and 5, and from the $\gamma$-grid outer-envelope and two-sided endpoint-coverage metrics reported for the primary conditional Cox–Breslow implementation in Table 1 of the main text and Web Table 8.

**Web Table 7.** Decision-regime and candidate-superior mirror simulations

(a) Decision-regime robustness and decision summary under covariate-dependent censoring

| Regime | True $\Delta R_T$ | True $\gamma^*$ cand | True $\gamma^*$ ref | $\gamma^*$ RMSE | Sep./direct ID width ratio | CAND | REF | DEFER |
|---|---|---|---|---|---|---|---|---|
| Candidate superior | -0.0144 | 0.38 | 0 | 0.051 | 3.74 | 0.355 | 0.000 | 0.645 |
| Reference superior | +0.0200 | 0 | 0.44 | 0.052 | 3.18 | 0.000 | 0.730 | 0.270 |
| Ambiguous | -0.0023 | <0.02 | 0 | 0.031 | 1.97 | 0.000 | 0.000 | 1.000 |
| Finite crossing | -0.1076 | 0.76 | 0 | 0.055 | 1.50 | 1.000 | 0.000 | 0.000 |
| No crossing within range | -0.1076 | 0.76 | 0 | – | 1.50 | 1.000 | 0.000 | 0.000 |

Notes: Decisions were evaluated at the prespecified sensitivity value $\gamma = 0.30$. CANDIDATE SUPPORTED, REFERENCE SUPPORTED, and DEFER are Monte Carlo frequencies over 1000 replications. CAND denotes CANDIDATE SUPPORTED; REF denotes REFERENCE SUPPORTED. Across the 10 regime-by-censoring conditions, $\gamma$-grid two-sided endpoint coverage ranged from 0.939 to 0.950. Robustness values were evaluated on the prespecified sensitivity grid with spacing 0.02. For the ambiguous regime, the first sign change occurred below the grid resolution, so $\gamma^*$ is reported as <0.02 rather than as zero. True $\gamma^*_{cand}$ and $\gamma^*_{ref}$ denote the population direction-specific robustness values defined in Proposition 3. A value of 0 indicates an empty certification set in that direction. For the ambiguous regime, $\gamma^*_{cand} < 0.02$ at the grid resolution. For the no-crossing-within-range regime, the underlying population $\gamma^*_{cand} = 0.76$, but the prespecified search range ended at 0.50; empirical robustness was therefore reported as >0.50, and robustness-value RMSE was not computed. The RMSE column refers to the relevant certification direction. The finite-crossing and no-crossing-within-range regimes use the same data-generating mechanism and differ only in the prespecified sensitivity search range.

(b) Decision frequencies by censoring mechanism at the prespecified decision sensitivity value $\gamma = 0.30$

| Regime | Censoring mechanism | CAND | REF | DEFER |
|---|---|---|---|---|
| Candidate superior | Independent | 0.419 | 0.000 | 0.581 |
| | Covariate-dependent | 0.355 | 0.000 | 0.645 |
| Reference superior | Independent | 0.000 | 0.771 | 0.229 |
| | Covariate-dependent | 0.000 | 0.730 | 0.270 |
| Ambiguous | Independent | 0.000 | 0.000 | 1.000 |
| | Covariate-dependent | 0.000 | 0.000 | 1.000 |
| Finite crossing | Independent | 1.000 | 0.000 | 0.000 |
| | Covariate-dependent | 1.000 | 0.000 | 0.000 |
| No crossing within range | Independent | 1.000 | 0.000 | 0.000 |
| | Covariate-dependent | 1.000 | 0.000 | 0.000 |

Notes: Decisions were evaluated at the prespecified sensitivity value $\gamma = 0.30$. Frequencies are Monte Carlo proportions over 1000 replications. Independent censoring used exponential censoring with marginal Kaplan–Meier estimation; covariate-dependent censoring used exponential censoring with Kaplan–Meier estimation within prespecified covariate strata. The finite-crossing and no-crossing-within-range regimes yielded CANDIDATE SUPPORTED in every replication under both censoring mechanisms, whereas the ambiguous regime yielded DEFER throughout.

(c) Candidate-superior mirror simulations

| $n_T$ | Scenario | $\gamma$ | CAND | Wrong REF | DEFER | True-contrast containment (per $\gamma$) |
|---|---|---|---|---|---|---|
| 60 | No shift | 0.2 | 0.151 | 0.000 | 0.849 | 0.981 |
| | | 0.3 | 0.101 | 0.000 | 0.899 | 0.987 |
| | Profile-varying shift | 0.2 | 0.135 | 0.000 | 0.865 | 0.981 |
| | | 0.3 | 0.093 | 0.000 | 0.907 | 0.988 |
| 80 | No shift | 0.2 | 0.155 | 0.000 | 0.845 | 0.980 |
| | | 0.3 | 0.112 | 0.000 | 0.888 | 0.983 |
| | Profile-varying shift | 0.2 | 0.162 | 0.000 | 0.838 | 0.982 |
| | | 0.3 | 0.116 | 0.000 | 0.884 | 0.988 |

Notes: Wrong REF denotes certification in favor of the reference predictor when the candidate predictor is truly superior. The final column reports per-$\gamma$ true-contrast

containment by the direct simultaneous band, defined as whether $\underline{L}_{\gamma} \leq \Delta R_T \leq \overline{U}_{\gamma}$; it is not $\gamma$-grid two-sided endpoint coverage. No wrong-direction certification occurred in any of the reported mirror cells. With 0 events in 1000 Monte Carlo replications, the exact one-sided 95% binomial upper confidence bound for the wrong-direction probability is ~0.0030.

**Web Table 8.** Factorial sensitivity of $\gamma$-grid outer-envelope coverage of the primary conditional Cox–Breslow implementation to source calibration size and censoring

| $n_S$ | 65% censoring | 50% censoring | 38% censoring |
|---|---|---|---|
| 114 | 0.859 [0.836, 0.879] (40 events) | 0.911 [0.892, 0.927] (57 events) | 0.916 [0.897, 0.932] (71 events) |
| 200 | 0.899 [0.879, 0.916] (69 events) | 0.915 [0.896, 0.931] (100 events) | 0.934 [0.917, 0.948] (124 events) |
| 300 | 0.914 [0.895, 0.930] (105 events) | 0.938 [0.921, 0.951] (150 events) | 0.947 [0.931, 0.959] (186 events) |
| 400 | 0.907 [0.887, 0.923] (139 events) | 0.947 [0.931, 0.959] (200 events) | 0.947 [0.931, 0.959] (248 events) |
| 516 | 0.920 [0.902, 0.935] (180 events) | 0.950 [0.935, 0.962] (258 events) | 0.940 [0.924, 0.953] (320 events) |

Notes: The factorial analysis used the same primary-implementation data-generating mechanism with $n_T = 80$ fixed. Only source calibration size ($n_S = 114$, 200, 300, 400 or 516) and censoring level (approximately 65%, 50%, or 38%) were varied. Each cell used 1,000 Monte Carlo replications and $B = 199$ bootstrap replicates. Entries are $\gamma$-grid outer-envelope coverage probabilities with Wilson 95% Monte Carlo confidence intervals. Mean realized censoring rates matched the targeted levels, and no fatal fitting failures occurred. Parentheses give the mean number of source events, $n_S \times$ (1 − mean realized censoring rate). The $n_S = 200, 300,$ and 400 cells were added after the original six under a protocol fixed before they were run. The data-generating mechanism has correctly specified outcome and censoring models. Censoring levels refer to the whole follow-up; the population proportions of source patients censored before $\tau$, whose status at $\tau$ is therefore unknown, are about 59%, 43%, and 31% at the 65%, 50%, and 38% levels. In the NSCLC calibration sample ($n = 114$; overall censoring 69%), 31–33 patients died before two years, 32–39 were known to survive two years, and 43–49 (38–43%) were censored before two years across the four configurations; estimated IPCW weights did not exceed 4.9, and the censoring floor was not activated. In the Rotterdam calibration sample (overall censoring 38%), 4–5% were

censored before five years. These comparisons describe information at $\tau$ only; event timing, covariate-dependent censoring, and the covariate distribution also differ between the simulation and the applications.

**Web Table 9.** NSCLC cohorts, sample composition, and analytic roles

| Cohort | Analytic role | Sample size | Observed events |
|---|---|---|---|
| CNUHH $D_{train}$ | Predictor development | 228 | 70 |
| CNUHH $D_{cal}$ | Source calibration | 114 | 35 |
| Stanford | Target | 60 | 22 |
| VA | Target | 80 | 32 |

Notes: The CNUHH source data comprised 400 patients with lung cancer and manually delineated tumor regions of interest; 57 patients with small-cell lung cancer and one patient with a recorded survival time of zero were excluded, leaving 342 patients with NSCLC and 105 observed deaths. For each locked configuration, an event-stratified split produced development and calibration subsets with the counts shown; counts were identical across configurations, whereas membership varied. The calibration subset was used to estimate the source conditional survival probability and the censoring model. Of 88 initially available VA cases, seven lacked sufficient harmonized metabolic tumor volume and one had failed PET-mask registration. Target survival outcomes were not supplied to the outcome-free analysis and were used only in the retrospective comparison of Web Table 14.

**Web Table 10.** Locked construction of the primary inductive SparsePCA candidate

(a) Common predictor construction and preprocessing

| Component | Locked specification |
|---|---|
| Imaging input | PET-only radiomics from the harmonized 40%-SUVmax MTV; 1,112 features |
| Target covariates used for primary candidate construction | No |
| Clinical variables for $q_0$ and $q_1$ | Age, sex, ever smoker, log pack-years, stage T, stage N, stage M |
| Feature screening | Source-priority screening using source $D_{train}$ only; candidate top-$k$ grid = 10, 20, 30, 40, 50 |
| SparsePCA component grid | 1, 2, 3, 5 |
| SparsePCA settings | alpha = 1.0; ridge_alpha = 0.01; method = lars; max_iter = 500; tol = $10^{-4}$ |
| Clinical Cox penalty grid | 0.1, 1, 10, 100 |
| Residual Cox | Pure L2 ridge; grid = 0.1, 1, 10, 100 |
| Combination parameter $\beta$ | Grid = 0.1, 0.25, 0.5, 1 |
| Candidate structure | Clinical Cox fitted first; ridge Cox on SparsePCA components with clinical risk as a fixed offset |
| Final linear predictor | $\eta_{clin} + \beta \times$ standardized residual imaging risk |
| Baseline hazard | Breslow estimator fitted using CNUHH $D_{train}$ |
| Prediction horizon | 730 days |
| Missing-value handling | Per-feature median learned in $D_{train}$ or the corresponding source fold-training set |
| Raw-feature scaling | Center by $D_{train}$ median; divide by population SD; clip to [−8,8] |
| Sparse-component scaling | Center by $D_{train}$ mean; divide by population SD; clip to [−12,12] |
| Residual-risk scaling | Center and scale using $D_{train}$ mean and SD |

Notes: All tuning, screening, and dimensionality-reduction steps for the primary candidate used source data only. The general feature-screening module can compute target-distribution similarity quantities, but these quantities were discarded by the locked source-priority branch and did not affect feature ranking or hyperparameter tuning.

(b) Seed-specific locked hyperparameters

| Seed | Selected PET features $k$ | SparsePCA components | Clinical Cox ridge | Residual Cox ridge | $\beta$ |
|---|---|---|---|---|---|
| 42 | 20 | 2 | 100 | 1 | 1.0 |
| 52 | 10 | 2 | 100 | 10 | 0.5 |
| 62 | 10 | 5 | 100 | 1 | 0.5 |
| 72 | 10 | 1 | 100 | 100 | 0.5 |

Notes: The four seeds determine the event-stratified source split, source cross-validation folds, stability ranking, SparsePCA random state, and consequently the locked feature and hyperparameter configurations. The same locked construction was used for the final Stanford and VA analyses.

**Web Table 11.** Summary of the complete primary NSCLC sensitivity grid

| Target | Seed | Direction of plug-in ID across grid | Decision across grid |
|---|---|---|---|
| Stanford | 42 | Positive at small $\gamma$, crosses 0 by $\gamma = 0.3$ | DEFER |
| | 52 | Positive throughout | DEFER |
| | 62 | Negative at small $\gamma$, crosses 0 by $\gamma = 0.5$ | DEFER |
| | 72 | Negative at small $\gamma$, crosses 0 by $\gamma = 0.5$ | DEFER |
| VA | 42 | Positive throughout | DEFER |
| | 52 | Positive throughout | DEFER |
| | 62 | Negative at small $\gamma$, crosses 0 by $\gamma = 0.5$ | DEFER |
| | 72 | Negative throughout | DEFER |

Notes: Each row summarizes five sensitivity values, $\gamma \in \{0, 0.1, 0.2, 0.3, 0.5\}$. All 40 primary evaluation cells resulted in DEFER because the simultaneous confidence envelope contained zero. Representative numerical endpoints are reported in Section 4.3 and Figure 3 of the main text and in Web Appendix B.4.

**Web Table 12.** NSCLC prediction-horizon sensitivity analysis

| Horizon | Target | Configurations | CAND | REF | DEFER | Unstable cells |
|---|---|---|---|---|---|---|
| 1 year | Stanford | 20 | 0 | 0 | 20 | 0 |
| | VA | 20 | 0 | 0 | 20 | 0 |
| 3 years | Stanford | 20 | 0 | 0 | 20 | 0 |
| | VA | 20 | 0 | 0 | 20 | 0 |

Notes: CAND denotes CANDIDATE SUPPORTED; REF denotes REFERENCE SUPPORTED. Each row summarizes four locked predictor configurations across five sensitivity values. All additional horizon-specific evaluations resulted in DEFER, and no cell was flagged as unstable with respect to censoring positivity. The detailed cell-level results are available in the reproducibility materials.

**Web Table 13.** NSCLC source-to-target overlap diagnostics in the deployment-summary space

| Target | Seed | Domain AUC | Outside source range | ESS/target | Max density ratio |
|---|---|---|---|---|---|
| Stanford | 42 | 0.7830 | 1.67% | 90.15% | 2.5904 |
| | 52 | 0.7699 | 0.00% | 90.01% | 2.4385 |
| | 62 | 0.7632 | 1.67% | 87.91% | 2.5263 |
| | 72 | 0.7458 | 1.67% | 91.73% | 2.2732 |
| VA | 42 | 0.7279 | 1.25% | 92.22% | 2.4359 |
| | 52 | 0.7286 | 0.00% | 92.51% | 2.2832 |
| | 62 | 0.7365 | 0.00% | 89.88% | 2.6547 |
| | 72 | 0.7357 | 0.00% | 92.57% | 2.2601 |

Notes: Diagnostics were computed in the deployment-summary space $Z = (q_0, q_1)$. Domain AUC quantifies source–target separability; outside-source-range is the proportion of target observations lying outside the rectangular source calibration range; ESS is the importance-weight effective sample size as a percentage of target sample size. These diagnostics assess gross empirical support deterioration but do not establish the transport-positivity assumption.

**Web Table 14.** Retrospective target-outcome evaluation of the primary NSCLC candidate

| Target | Seed | ΔBrier | 95% CI | Δ C-index | 95% CI |
|---|---|---|---|---|---|
| Stanford | 42 | +0.0066 | [-0.0205, +0.0368] | +0.0274 | [-0.0368, +0.0859] |
| | 52 | -0.0078 | [-0.0322, +0.0101] | +0.0162 | [-0.0259, +0.0606] |
| | 62 | -0.0220 | [-0.0508, +0.0039] | +0.0424 | [-0.0312, +0.1134] |
| | 72 | -0.0071 | [-0.0293, +0.0123] | +0.0224 | [-0.0298, +0.0749] |
| VA | 42 | +0.0440 | [+0.0059, +0.0886] | -0.0048 | [-0.0671, +0.0549] |
| | 52 | +0.0153 | [-0.0068, +0.0459] | +0.0086 | [-0.0421, +0.0610] |
| | 62 | +0.0154 | [-0.0025, +0.0344] | -0.0365 | [-0.0948, +0.0215] |
| | 72 | +0.0202 | [-0.0019, +0.0508] | -0.0161 | [-0.0677, +0.0333] |

Notes: Contrasts are candidate minus reference. For Brier risk, negative values favor the candidate; for the C-index, positive values favor the candidate. These target-outcome analyses were performed separately from the finalized outcome-free decision procedure and are reported only as retrospective descriptive comparisons. They were not used to select the sensitivity parameter, construct the identified sets, or determine CANDIDATE/REFERENCE/DEFER decisions.

**Web Table 15.** Complete Rotterdam-to-GBSG results across locked configurations and sensitivity values

| Seed | $\gamma$ | Direct identified set | Separate identified set | Direct simultaneous envelope | Separate simultaneous envelope | Direct decision | Separate decision |
|---|---|---|---|---|---|---|---|
| 42 | 0.0 | [-0.0255, -0.0255] | [-0.0255, -0.0255] | [-0.0395, -0.0115] | [-0.0406, -0.0105] | CAND | CAND |
| | 0.1 | [-0.0307, -0.0202] | [-0.0348, -0.0161] | [-0.0442, -0.0058] | [-0.0491, -0.0002] | CAND | CAND |
| | 0.2 | [-0.0357, -0.0147] | [-0.0438, -0.0065] | [-0.0487, +0.0001] | [-0.0574, +0.0101] | DEFER | DEFER |
| | 0.3 | [-0.0404, -0.0091] | [-0.0527, +0.0032] | [-0.0530, +0.0061] | [-0.0655, +0.0206] | DEFER | DEFER |
| | 0.5 | [-0.0493, +0.0026] | [-0.0697, +0.0229] | [-0.0608, +0.0184] | [-0.0811, +0.0417] | DEFER | DEFER |
| 52 | 0.0 | [-0.0202, -0.0202] | [-0.0202, -0.0202] | [-0.0352, -0.0053] | [-0.0359, -0.0045] | CAND | CAND |
| | 0.1 | [-0.0258, -0.0144] | [-0.0306, -0.0096] | [-0.0403, +0.0009] | [-0.0455, +0.0068] | DEFER | DEFER |
| | 0.2 | [-0.0313, -0.0085] | [-0.0408, +0.0010] | [-0.0453, +0.0072] | [-0.0549, +0.0182] | DEFER | DEFER |
| | 0.3 | [-0.0365, -0.0024] | [-0.0508, +0.0118] | [-0.0500, +0.0136] | [-0.0641, +0.0298] | DEFER | DEFER |
| | 0.5 | [-0.0463, +0.0100] | [-0.0700, +0.0336] | [-0.0588, +0.0265] | [-0.0819, +0.0529] | DEFER | DEFER |
| 62 | 0.0 | [-0.0264, -0.0264] | [-0.0264, -0.0264] | [-0.0394, -0.0134] | [-0.0402, -0.0126] | CAND | CAND |
| | 0.1 | [-0.0315, -0.0211] | [-0.0364, -0.0162] | [-0.0441, -0.0076] | [-0.0495, -0.0016] | CAND | CAND |
| | 0.2 | [-0.0365, -0.0156] | [-0.0463, -0.0058] | [-0.0486, -0.0017] | [-0.0586, +0.0096] | **CAND** | **DEFER** |
| | 0.3 | [-0.0412, -0.0099] | [-0.0559, +0.0047] | [-0.0529, +0.0043] | [-0.0676, +0.0208] | DEFER | DEFER |
| | 0.5 | [-0.0501, +0.0017] | [-0.0744, +0.0260] | [-0.0608, +0.0165] | [-0.0848, +0.0434] | DEFER | DEFER |
| 72 | 0.0 | [-0.0175, -0.0175] | [-0.0175, -0.0175] | [-0.0329, -0.0020] | [-0.0336, -0.0013] | CAND | CAND |
| | 0.1 | [-0.0232, -0.0116] | [-0.0289, -0.0059] | [-0.0381, +0.0043] | [-0.0441, +0.0111] | DEFER | DEFER |
| | 0.2 | [-0.0287, -0.0055] | [-0.0401, +0.0058] | [-0.0432, +0.0107] | [-0.0545, +0.0237] | DEFER | DEFER |
| | 0.3 | [-0.0341, +0.0006] | [-0.0511, +0.0176] | [-0.0481, +0.0172] | [-0.0646, +0.0363] | DEFER | DEFER |
| | 0.5 | [-0.0441, +0.0132] | [-0.0722, +0.0413] | [-0.0571, +0.0302] | [-0.0842, +0.0615] | DEFER | DEFER |

Notes: CAND denotes CANDIDATE SUPPORTED. Negative Brier-risk contrasts favor the candidate. Decisions are based on the simultaneous confidence envelopes, not on the plug-in

identified sets alone. The four seeds represent distinct locked predictor configurations and are not repeated trials or components of a voting rule. At seed 62 and $\gamma = 0.2$, direct identification yielded CANDIDATE SUPPORTED whereas separate-risk subtraction yielded DEFER. The predictor contrast was motivated by established prognostic knowledge and specified without using GBSG target outcomes. Source nuisance estimation used the same conditional Cox–Breslow censoring model and IPCW-weighted logistic outcome model as in the main analysis. Estimated population robustness values for candidate superiority were 0.455, 0.335, 0.470, and 0.285 under direct identification and 0.265, 0.190, 0.250, and 0.150 under separate subtraction (seeds 42, 52, 62, 72; grid step 0.005).

**Web Table 16.** Plug-in versus one-step endpoint estimators in simulation (1,000 replications per cell; $B = 199$)

| Scale | Law | Estimator | Bias ($\times 10^{-3}$) range over $\gamma$ and end points | SD ($\times 10^{-3}$) | RMSE ($\times 10^{-3}$) | Outer coverage [Wilson 95% CI] | Mean envelope width at $\gamma$ = 0.2 ($\times 10^{-3}$) |
|---|---|---|---|---|---|---|---|
| NSCLC | A (correct) | Plug-in | 1.9 to 12.2 | 26.9-29.8 | 27.0-32.2 | 0.863 [0.840, 0.883] | 130.3 |
| | | One-step | 1.8 to 13.8 | 28.5-32.0 | 28.5-34.8 | 0.844 [0.820, 0.865] | 134.9 |
| | B (misspecified) | Plug-in | 3.8 to 13.5 | 27.4-30.1 | 27.6-33.0 | 0.860 [0.837, 0.880] | 131.7 |
| | | One-step | 2.8 to 13.9 | 29.3-32.0 | 29.4-34.9 | 0.847 [0.823, 0.868] | 137.0 |
| Rotterdam-GBSG | A (correct) | Plug-in | -0.3 to 1.0 | 8.2-8.9 | 8.2-8.9 | 0.942 [0.926, 0.955] | 54.4 |
| | | One-step | -0.3 to 1.1 | 8.5-9.2 | 8.5-9.2 | 0.949 [0.934, 0.961] | 55.9 |
| | B (misspecified) | Plug-in | 1.2 to 2.2 | 7.8-8.2 | 7.9-8.4 | 0.957 [0.943, 0.968] | 54.2 |
| | | One-step | 0.1 to 1.2 | 8.1-8.6 | 8.2-8.6 | 0.957 [0.943, 0.968] | 55.6 |

Notes: Law A is the correctly specified source outcome law; law B adds the omitted term $2.5(q_1 - 0.55)^2$ to the logit of the source outcome law, while the working model remains logistic in standardized $(q_0, q_1)$. Bias, SD, and RMSE are ranges over $\gamma \in \{0, 0.1, 0.2, 0.3, 0.5\}$ and both endpoints. Population limits of the bias were 0 for both estimators under law A, and 1.08–1.15 (plug-in) and −0.02 to 0.02 (one-step) under law B ($\times 10^{-3}$), omputed by fitting the working model to the true source law on 2,000,000 source draws and evaluating both limits on 2,000,000 target draws. Outer coverage is simultaneous over the grid, with nominal level 0.95. No fatal fits occurred; the density-ratio cap was never

activated at point estimates; the censoring floor was activated at the point estimate in 2 of 1,000 law-B NSCLC replications.

**Web Table 17.** Plug-in versus one-step endpoint estimates in the NSCLC and Rotterdam–GBSG applications at $\gamma = 0.2$

| Setting | Seed | Plug-in identified set | One-step identified set | Plug-in envelope | One-step envelope | Decisions (plug-in/one-step) |
|---|---|---|---|---|---|---|
| NSCLC–Stanford | 42 | [+0.0021, +0.0147] | [-0.0058, +0.0058] | [-0.0237, +0.0455] | [-0.0323, +0.0350] | DEFER/DEFER |
| | 52 | [+0.0105, +0.0191] | [+0.0079, +0.0151] | [-0.0115, +0.0421] | [-0.0146, +0.0380] | DEFER/DEFER |
| | 62 | [-0.0123, -0.0032] | [-0.0115, -0.0039] | [-0.0459, +0.0299] | [-0.0424, +0.0269] | DEFER/DEFER |
| | 72 | [-0.0110, -0.0042] | [-0.0167, -0.0121] | [-0.0292, +0.0160] | [-0.0329, +0.0052] | DEFER/DEFER |
| NSCLC–VA | 42 | [+0.0248, +0.0410] | [+0.0173, +0.0329] | [-0.0267, +0.0958] | [-0.0318, +0.0841] | DEFER/DEFER |
| | 52 | [+0.0187, +0.0312] | [+0.0164, +0.0278] | [-0.0215, +0.0725] | [-0.0227, +0.0672] | DEFER/DEFER |
| | 62 | [-0.0156, -0.0067] | [-0.0144, -0.0065] | [-0.0479, +0.0270] | [-0.0442, +0.0253] | DEFER/DEFER |
| | 72 | [-0.0164, -0.0087] | [-0.0229, -0.0176] | [-0.0386, +0.0162] | [-0.0436, +0.0047] | DEFER/DEFER |
| Rotterdam–GBSG | 42 | [-0.0357, -0.0147] | [-0.0345, -0.0135] | [-0.0487, +0.0001] | [-0.0478, +0.0016] | DEFER/DEFER |
| | 52 | [-0.0313, -0.0085] | [-0.0306, -0.0079] | [-0.0453, +0.0072] | [-0.0448, +0.0080] | DEFER/DEFER |
| | 62 | [-0.0365, -0.0156] | [-0.0362, -0.0151] | [-0.0486, -0.0017] | [-0.0486, -0.0010] | CAND/CAND |
| | 72 | [-0.0287, -0.0055] | [-0.0301, -0.0069] | [-0.0432, +0.0107] | [-0.0441, +0.0088] | DEFER/DEFER |

Notes: CAND denotes CANDIDATE SUPPORTED. The one-step estimator was computed from the same bootstrap draws ($B = 999$) and critical-value grid as the reported plug-in analysis. The plug-in columns therefore reproduce Figures 3 and 4 and Web Table 15 exactly. Decisions agreed in all 60 configuration-by-sensitivity cells; the full grid is in the reproducibility archive. Target outcomes were not read.

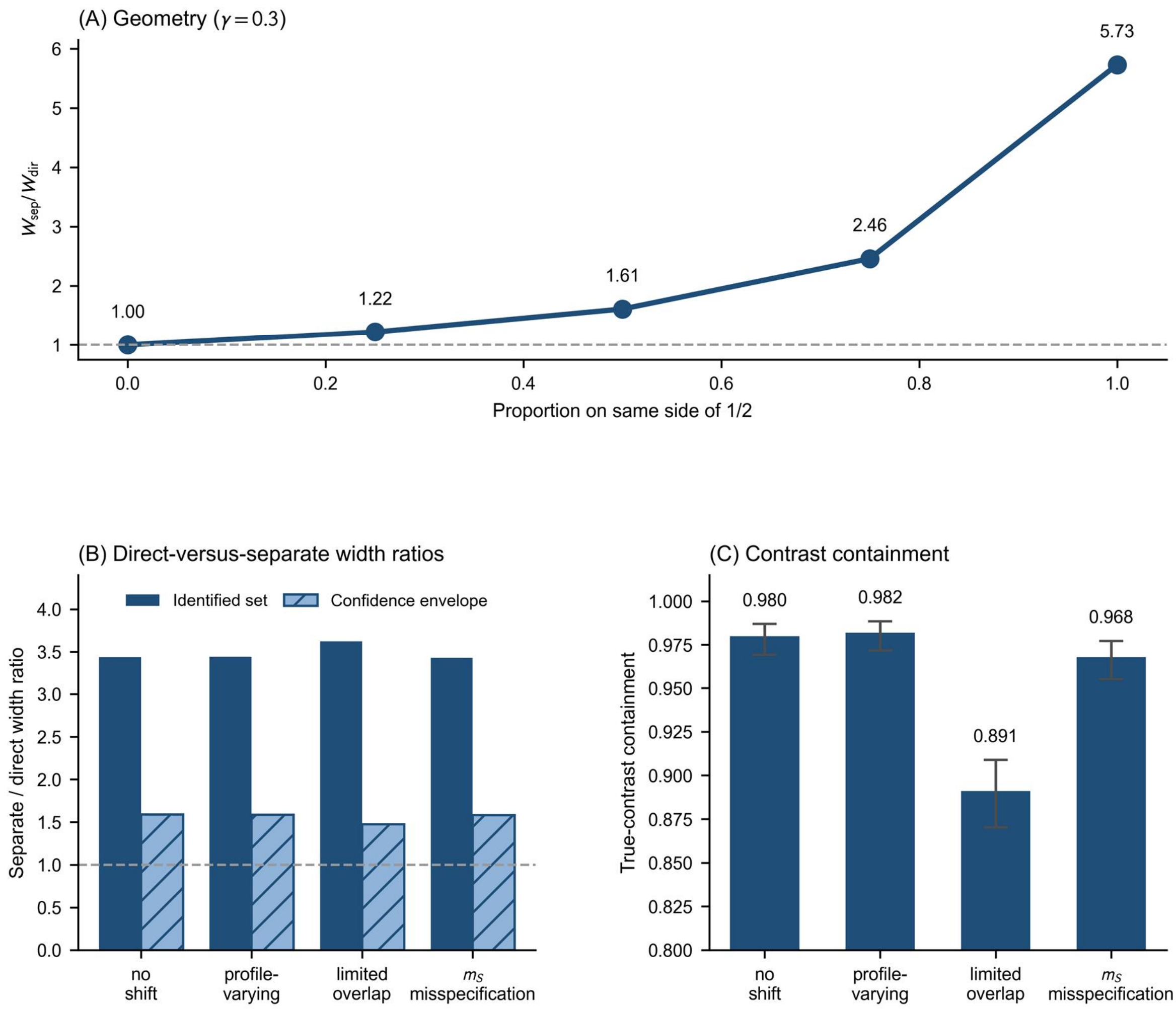

**Web Figure 1.** Analytic identification geometry and application-scale comparator performance. (A) Separate-to-direct identified-set width ratio at $\gamma = 0.3$ as a function of the proportion of target profiles with both predictions on the same side of one half, based on a single numerical sample of 300,000 target profiles; this panel is not a Monte Carlo coverage study. (B) Application-scale separate-to-direct width ratios for $n_T = 80$ at $\gamma = 0.2$, shown for the sharp identified set and the simultaneous confidence envelope (marginal Kaplan–Meier comparator). (C) Per-$\gamma$ true-contrast containment by the direct simultaneous band for the same scenarios; error bars are Monte Carlo 95% confidence intervals. Containment of the single true contrast is not endpoint coverage. In panels B and C, $m_S$ misspecification denotes source outcome-model misspecification.

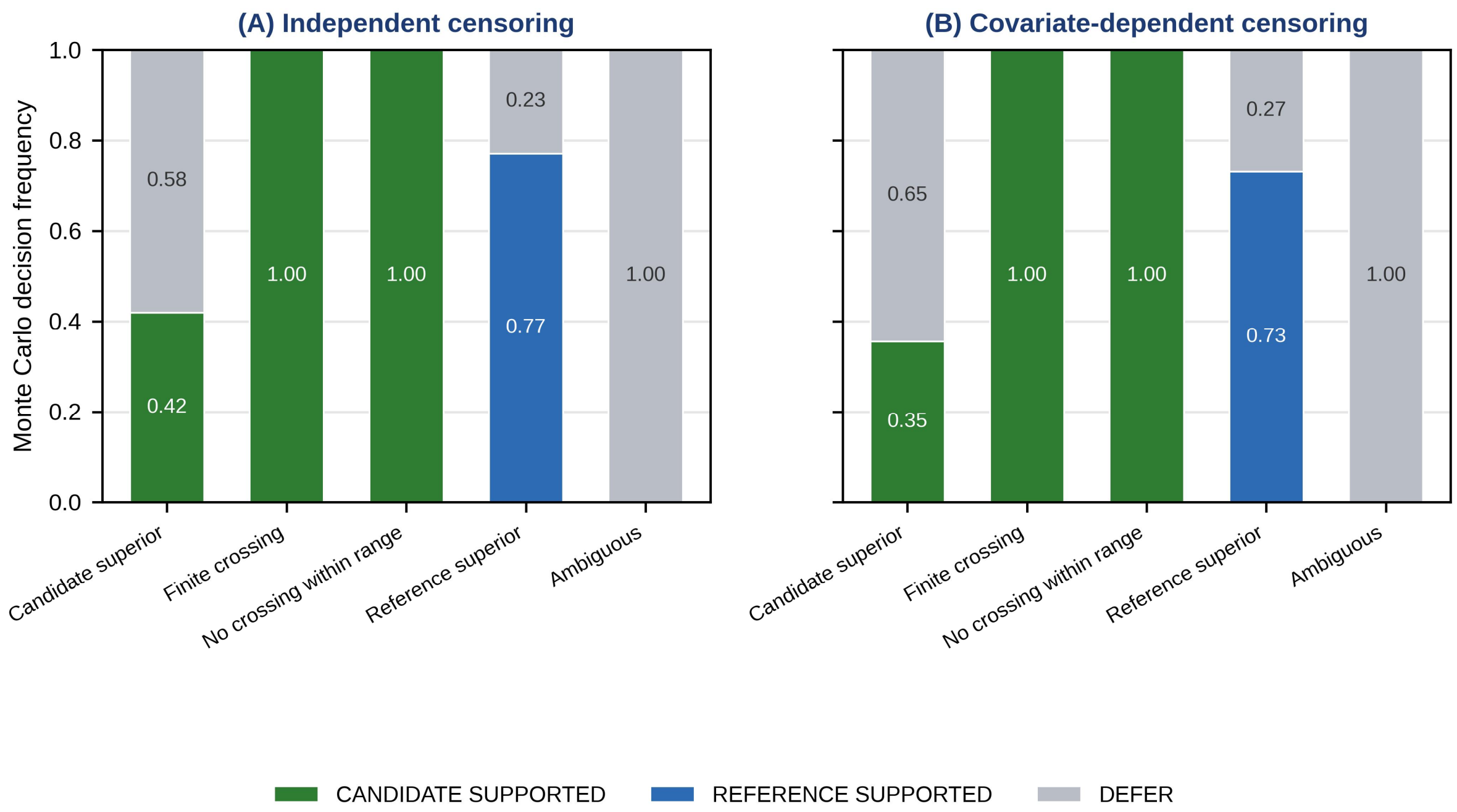

**Web Figure 2.** Decision frequencies across prespecified simulation regimes

Monte Carlo frequencies of CANDIDATE SUPPORTED, REFERENCE SUPPORTED, and DEFER are shown across five prespecified decision regimes under (A) independent censoring and (B) covariate-dependent censoring. Decisions were based on the simultaneous confidence-envelope rule. The candidate-superior regime yielded a mixture of CANDIDATE SUPPORTED and DEFER, the finite-crossing and no-crossing-within-range regimes yielded CANDIDATE SUPPORTED in all replications, the reference-superior regime yielded predominantly REFERENCE SUPPORTED, and the ambiguous regime yielded DEFER throughout. Frequencies summarize repeated Monte Carlo samples and should not be interpreted as voting across predictor configurations.

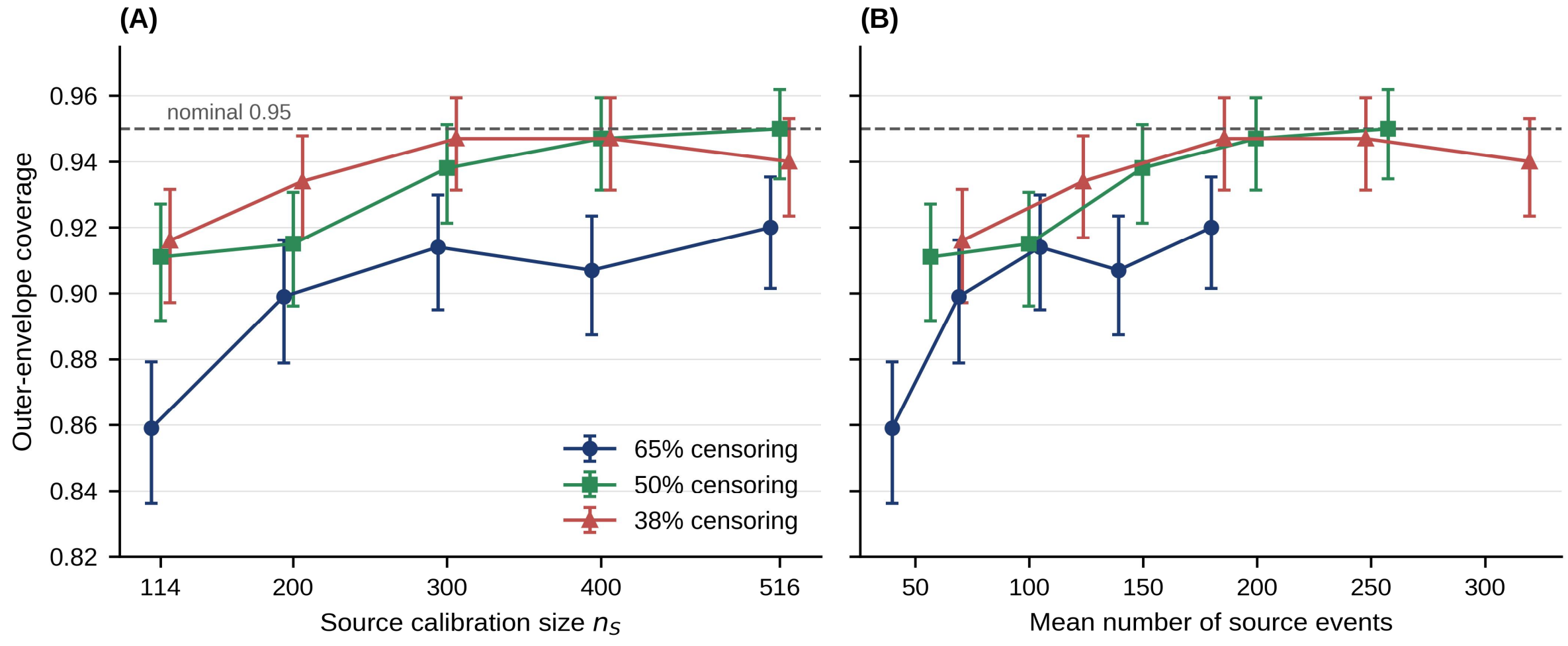

**Web Figure 3.** Simultaneous $\gamma$-grid outer-envelope coverage of the primary conditional Cox–Breslow implementation across the 15 factorial cells, by (A) source calibration size and (B) mean number of source events. Lines connect cells with the same censoring level; in (B), points on each line correspond from left to right to $n_S = 114, 200, 300, 400,$ and 516. Vertical bars are Wilson 95% Monte Carlo intervals, and the dashed line marks the nominal 0.95. $n_T = 80$; 1,000 replications per cell; $B = 199$. Points in (A) are slightly offset horizontally for legibility.

## References for the Web Appendix